\documentclass{emulateapj}
\usepackage{xspace}
\newcommand{\zlens}{\ensuremath{z_{\rm lens}}\xspace}
\newcommand{\rvir}{\ensuremath{r_{\rm vir}}\xspace}
\newcommand{\ebv}{\ensuremath{E(\bv)}\xspace}

\newcommand\avg[1]{\langle{#1}\rangle}
\slugcomment{Accepted for publication in the Astrophysical Journal} 
\shorttitle{Gravitational Lens Environments}
\shortauthors{Williams et al.}

\begin{document}
\title{First Results from a Photometric Survey of Strong Gravitational
  Lens Environments}

\author{Kurtis A. Williams}
\email{kurtis@as.arizona.edu}
\author{Ivelina Momcheva}
\email{imomcheva@as.arizona.edu}
\affil{Steward Observatory \\ 933 N. Cherry Ave., Tucson, AZ 85721}
\author{Charles R. Keeton}
\email{keeton@physics.rutgers.edu}
\affil{Department of Physics and Astronomy \\ Rutgers University \\
  136 Frelinghuysen Rd., Piscataway, NJ 08854}
\author{Ann I. Zabludoff}
\email{aiz@as.arizona.edu}
\affil{Steward Obs. \\ 933 N. Cherry Ave., Tucson, AZ 85721}
\and
\author{Joseph Leh\'{a}r\altaffilmark{1}}
\email{jlehar@combinatorx.com}
\affil{Harvard-Smithsonian Center for Astrophysics \\ 60 Garden Street
  \\ Cambridge, MA 02138} 
\altaffiltext{1}{Present address: CombinatoRx,
    Inc., 650 Albany Street, Boston, MA 02118}

\begin{abstract}
Many strong gravitational lenses lie in complex environments, such as
poor groups of galaxies, that significantly bias conclusions from lens
analyses.  We are undertaking a photometric survey of all known
galaxy-mass strong lenses to characterize their environments and
include them in careful lens modeling, and to build a large, uniform
sample of galaxy groups at intermediate redshifts for evolutionary
studies.  In this paper we present wide-field photometry of the
environments of twelve lens systems with $0.24\leq \zlens\leq 0.5$.
Using a red-sequence identifying technique, we find that eight of the
twelve lenses lie in groups, and that ten group-like structures are
projected along the line of sight towards seven of these lenses.
Follow-up spectroscopy of a subset of these fields confirms these
results.  For lenses in groups, the group centroid position is
consistent with the direction of the external tidal shear required by
lens models.  Lens galaxies are not all super-$L_*$ ellipticals; the
median lens luminosity is $\lesssim L_*$, and the distribution of lens
luminosities extends 3 magnitudes below $L_*$ (in agreement with
theoretical models).  Only two of the lenses in groups are the
brightest group galaxy, in qualitative agreement with theoretical
predictions.  As in the local Universe, the highest
velocity-dispersion ($\sigma$) groups contain a brightest member
spatially coincident with the group centroid, whereas lower-$\sigma$
groups tend to have an offset brightest group galaxy.  This suggests
that higher-$\sigma$ groups are more dynamically relaxed than
lower-$\sigma$ groups and that at least some evolved groups exist by
$z\sim 0.5$.
\end{abstract}

\section{Introduction}
Strong gravitational lensing of active galactic nuclei by intervening
galaxies is a versatile tool for studying many astrophysical issues.
Measurements of the time delay in brightness fluctuations between
different images, when combined with an assumed mass model, give a
measurement of the Hubble constant independent of the canonical
distance ladder
\citep[e.g.,][]{Refsdal1964,Vanderriest1989,Roberts1991,Blandford1992,
Keeton1997,Courbin2002,Kochanek2004b,Schechter2005}.  Gravitational
lensing statistics can be used to constrain the density and the
equation of state of dark energy
\citep[e.g.,][]{Turner1990,Kochanek1996a,Cooray1999,Chae2003,Mitchell2005}.
Strong lensing also allows the determination of the mass, shape and
evolution of dark-matter halos
\citep[e.g.,][]{Kochanek1991,Keeton1998,Treu2002,Treu2002b,Rusin2003,Rusin2005}.
Anomalies in the flux ratios of individual lenses are the basis for
studies of substructure in dark matter halos
\citep[e.g.,][]{Mao1998,Metcalf2001,Dalal2002,Metcalf2004},
measurements of the size of AGN accretion disk sizes
\citep[e.g.,][]{Rauch1991,Wyithe2000,Yonehara2001,Kochanek2004} and
the size of AGN broad-line regions
\citep[e.g.,][]{Abajas2002,Lewis2004,Richards2004}.

These studies are all sensitive to the mass model used to calculate
the lensing potential.  Understanding whether the lens potential is
significantly affected by a complex environment is crucial for
obtaining accurate models and avoiding significant biases
\citep{Keeton2004}.  Simple arguments suggest that at least 25\% of
lens galaxies should lie in bound groups or clusters of galaxies
\citep{Keeton2000}, but most lens environments have not yet been
studied.  A few gravitational lenses are already known to lie in galaxy
groups \citep[e.g.,][]{Kundic1997,Tonry1998,Tonry1999,Fassnacht2006}
and clusters \citep[e.g.,][]{Young1981,Fischer1998,Kneib2000}.
Extended X-ray emission, indicative of more massive, dynamically
evolved groups has been detected around some lenses \citep[PG1115+080
and CLASS B1422+231;][]{Grant2004}, but not around others \citep[CLASS
B1600+434 and CLASS B1608+656;][]{Dai2005}.

Attempts have been made to correlate lens models with lens
environments, with mixed results
\citep[e.g.,][]{Hogg1994,Schechter1997,Keeton1997,Kundic1997,Kundic1997b,Lehar2000}.
However, these studies typically examine a region only $\sim
30\arcsec$ ($\sim 130$ kpc at $z=0.3$) around the lens, and group
virial radii \citep[$\sim 500h^{-1}$ kpc;][]{Zabludoff1998} can extend
out to a few arcminutes.  Most recently, Momcheva et al.~\citep[2006;
hereafter][]{Momcheva2005} find that six of eight lenses in that
spectroscopic survey lie in groups.  Some of those lenses were
targeted because a group was previously known or suspected, so it is
not yet clear whether \citet{Momcheva2005}'s high group fraction is
representative of most lenses.  A larger, less-biased survey of lens
environments is clearly required.

As gravitational lensing involves an integration of mass along the
line-of-sight, one must also consider the possible effect of
line-of-sight galaxy groups, both foreground and background.  If the
impact parameter is small and the mass large, an interloping group can
have a significant impact ($\gtrsim 0.05$ in the normalized shear and
convergence terms) in lensing models \citep{Momcheva2005}, although
theoretical estimates of whether this effect is common yield
conflicting results
\citep[e.g.,][]{Seljak1994,BarKana1996,Keeton1997b,Premadi2004}.
Observational studies find evidence of line-of-sight groups toward a
few gravitational lenses, such as B0712+472 \citep{Fassnacht2002}, MG
J1131+0445 \citep{Tonry2000}, and B1608+656 \citep{Fassnacht2006}.  In
\citet{Momcheva2005} we present spectroscopic identification of a
massive group behind MG J0751+2716 that significantly impacts that
lens model.  A detailed census of line-of-sight groups is clearly
needed to understand whether or not they are common, significant
perturbers of lens models.

If many lenses do lie in groups, a comprehensive study of lens galaxy
environments will provide a unique sample of galaxy groups with
redshifts out to $z\sim 1$, the range of lens galaxy redshifts.  The
group environment is an important laboratory for studying galaxy
evolution, as groups are common environments for galaxies and are
relatively simple systems in which the range of factors impacting
galaxy evolution (primarily galaxy-galaxy interactions) is much
narrower than in the complex, hot, dense cluster environment
\citep{Zabludoff1998}.  There is a small but growing number of surveys
for groups at intermediate redshifts
\citep{Carlberg2001,Wilman2005,Gerke2005}.  These searches are
challenging, because poor groups are difficult to find given their low
projected surface densities and faint X-ray luminosities.  The large,
homogeneous sample of groups obtained from a lens
survey would, when compared with local samples, permit direct
observation of group evolution.

One measure of group evolution may be the properties of the brightest
group galaxy (BGG).  In nearby, high velocity dispersion
($\sigma\gtrsim 300\,{\rm km\, s}^{-1}$) groups, the BGG is typically
a giant, central elliptical \citep{Zabludoff1998,van den Bosch2005}.
In contrast, low-$\sigma$ groups at $z\approx 0$ rarely have a
dominant, bright member at their centers and tend to be
elliptical-poor.  These results suggest that higher-$\sigma$ groups
are generally more dynamically evolved than their low-$\sigma$
counterparts.  It is unknown whether this dichotomy persists out to
higher redshifts or whether group evolution is sufficiently rapid to
make distant groups with central giant ellipticals rare.  A sample of
poor groups at higher redshift may permit this question to be
addressed for the first time.

Because lensing selects on the basis of mass
\citep[e.g.,][]{Fukugita1991}, one might first expect that lens
galaxies in groups would also be the BGGs. However, the sheer number
of dwarf galaxies in a rich environment can offset the larger cross
section of the massive BGG \citep{Keeton2000}, leading to the
prediction that fewer than $\sim$50\% of lens galaxies are BGGs
\citep{Oguri2005}.  Lens theory further suggests that lens galaxies
span a large range of luminosities, with a typical luminosity around
$L_*$ \citep[e.g.,][]{Kochanek2000}.  The data are now in hand to
determine directly the distribution of lens galaxy luminosities and
the fraction of lens galaxies that are also BGGs.

In order to address these issues, we have undertaken a photometric
survey of $\sim 80$ strong gravitational lenses, nearly the complete
sample of lens systems known as of early 2004, with a significant
fraction of these ($\sim 1/3$) targeted for follow-up spectroscopy.
The goals of our survey are to characterize the local environment of
each lens galaxy out to large radii ($\sim 15$ arcminutes), to identify
any line-of-sight structures toward each lens system, to determine the
impact of the local environment and the interloping systems on the
lens model, and to compile a large catalog of galaxy groups at
intermediate redshifts useful for studying group evolution from
$z\approx 1$ to the present.

In \citet{Keeton2004}, the theoretical underpinnings of the biases
introduced by over-dense environments at the lens redshift are
discussed in detail.  In this paper we present the first results of
our photometric survey, focusing on the fields of twelve lenses.
Follow-up spectroscopy of eight of these lenses and detailed analyses
of the shear and convergence introduced by both the immediate lens
environment and any interloping structures are presented in
\citet{Momcheva2005}.  In \S 2 of this paper we describe
the lens sample, observations, and data reduction techniques.  In
\S 3 we discuss the red-sequence finding algorithm
we use to detect candidate groups in our photometric catalog,
including simulations of its effectiveness and comparisons to the
spectroscopic data.  In \S 4 we employ our sample of
detected groups to address the issues raised above.  Throughout this
paper we assume a standard cosmology of $H_0=70$ km s$^{-1}$ Mpc
$^{-1}$, $\Omega_m=0.3$, and $\Omega_\Lambda = 0.7$.  Unless
explicitly stated otherwise, all physical quantities from referenced
papers have been adapted to this cosmology.

\section{Observations\label{sec.obs}}

\subsection{Sample Selection}

We select our sample of twelve lens systems from the sample of all
known lens systems on the basis of redshift ($\zlens \leq 0.5$) and
foreground reddening ($\ebv\leq 0.2$).  We also select roughly equal
numbers of quadruple- and double-image lenses, as well as a handful of
``other'' image types (two objects with Einstein rings and one system
with two lens planes).  Imaging and analysis of the remaining lens
systems will be presented in a future paper.  The current lens sample
is given in Table~\ref{tab.sample}.

\begin{deluxetable*}{lccccccccccccc}
\tabletypesize{\scriptsize}
\tablewidth{0pt}
\tablecolumns{14}
\tablecaption{Selected lens systems\label{tab.sample}}
\tablehead{
  \colhead{Lens Name} & \colhead{$N_{\rm im}$\tablenotemark{a}} &
  \colhead{\zlens\tablenotemark{b}} & $\ebv$ & 
  \multicolumn{3}{c}{Obs. Night\tablenotemark{c}} & 
  \multicolumn{3}{c}{Exp. time (s)} & 
  \multicolumn{3}{c}{Calib. Night\tablenotemark{d}} & 
  \colhead{References\tablenotemark{e}}\\
  & & & &
  \colhead{$V$} & \colhead{$R$} & \colhead{$I$} & 
  \colhead{$V$} & \colhead{$R$} & \colhead{$I$} & 
  \colhead{$V$} & \colhead{$R$} & \colhead{$I$} &
}
\startdata
{CLASS B0712+472} & 4 & 0.414 & 0.113 & \nodata & K1,K4 & K3 & \nodata & 1800 & 2700 & \nodata & \nodata & \nodata & 1,2\\
{MG J0751+2716} & R & 0.350 & 0.034 & \nodata & K2 & K3 & \nodata & 1800 & 2700 & \nodata & \nodata & \nodata & 3,4\\
{FBQS J0951+2635}& 2 & 0.24\tablenotemark{f} & 0.022 & K2,K3 & \nodata & K3,K4 & 2700 & \nodata & 2700 & \nodata & \nodata & \nodata & 5,6\\
{BRI 0952-0115} & 2 & 0.41  & 0.063 & \nodata & C2 & C3 & \nodata & 1800 & 1800 & \nodata & C5 & C5 & 6,7\\
{PG 1115+080}   & 4 & 0.310 & 0.041 & C3 & \nodata & C1 & 1800 & \nodata & 2400 & SDSS    & \nodata & \nodata & 8,9,10\\
{1RXS J113155.4-123155} & 4 & 0.295 & 0.035 & C7 & \nodata & C7,C8 & 2400 & \nodata & 2400 & \nodata & \nodata & \nodata & 11\\
{CLASS B1422+231} & 4 & 0.339 & 0.048 & K1 & \nodata & K3 & 1800 & \nodata & 2700 & \nodata & \nodata & \nodata & 9,10,12\\
{CLASS B1600+434} & 2 & 0.414 & 0.013 & \nodata & K2 & K4 & \nodata & 1800 & 1800 & \nodata & \nodata & K8 & 2,13\\
{MG J1654+1346} & R & 0.254 & 0.061 & C2 & \nodata & C1,C2 & 2700 & \nodata & 4500 & K8 & \nodata & \nodata & 14\\
{PMN J2004-1349}& 2 & $\lesssim 0.36$ & 0.202 & \nodata & C2 & C3 & \nodata & 1800 & 1800 & \nodata & C9 & C9 & 15,16\\
{CLASS B2114+022}& 2+2? & 0.316, 0.588 & 0.072 & \nodata & C2 & C2 & \nodata & 1800 & 1800 & \nodata & C9 & C9 & 17,18\\
{HE 2149-2745} & 2 & 0.495 & 0.032 & \nodata & C1 & C1 & \nodata & 1800 & 1800 & \nodata & \nodata & \nodata & 19,20\\ 
\enddata
\tablenotetext{a}{Number of images; R indicates an Einstein Ring}
\tablenotetext{b}{From literature or CASTLES}
\tablenotetext{c}{See Table~\ref{tab.obs}}
\tablenotetext{d}{Where necessary}
\tablenotetext{e}{References for lens discovery and lens galaxy
  redshift}
\tablenotetext{f}{Photometric Redshift}
\tablerefs{ (1) \citealt{Jackson1998}, (2) \citealt{Fassnacht1998},
(3) \citealt{Lehar1997}, (4) \citealt{Tonry1999}, (5) \citealt{Schechter1998}, (6) \citealt{Kochanek2000},
(7) \citealt{McMahon1992}, (8)\citealt{Weymann1980}, (9) \citealt{Kundic1997}, (10) \citealt{Tonry1998}
(11) \citealt{Sluse2003}, (12) \citealt{Patnaik1992}, (13) \citealt{Jackson1995}, (14) \citealt{Langston1989},
(15) \citealt{Winn2001}, (16) \citealt{Winn2003}, (17) \citealt{King1996a}, (18) \citealt{Augusto2001},
(19) \citealt{Wisotzki1996}, (20) \citealt{Burud2002}}

\end{deluxetable*}

Of these twelve lens systems, three have previously published
spectroscopically confirmed groups at the lens redshift: {MG
J0751+2716} \citep[hereafter MG0751;][]{Tonry1999}, {PG
1115+080} \citep[hereafter PG1115;][]{Kundic1997}, and {CLASS
B1422+231} \citep[hereafter B1422;][]{Kundic1997}.  {B0712+472}
(hereafter B0712) has a spectroscopic foreground group at $z\approx
0.29$ but no known group surrounding the lens \citep{Fassnacht2002}.
Two other lens systems have either an over-density of galaxies
projected around the lens \citep[{MG J1654+1346}, hereafter
MG1654;][]{Langston1989} or a group identified by photometric
redshifts \citep[{HE 2149-2745}, hereafter
HE2149;][]{Lopez1998,Faure2004}. Two more lens systems are best
characterized by models including external shear, suggesting the
presence of an additional mass component: {BRI 0952-0115}
\citep[hereafter BRI0952;][]{Lehar2000} and {1RXS
J113155.4-123155} \citep[hereafter RXJ1131;][]{Sluse2003}.  For the
remaining four lenses, {FBQS J0951+2635} (hereafter FBQS0951),
{CLASS B1600+434} (hereafter B1600), {PMN J2004-1349}
(hereafter PMN2004), and {CLASS B2114+022} (hereafter B2114),
no evidence of a complex environment has been found previously.
 
It is not yet clear how representative these twelve systems are of all
lens environments.  This sample includes all eight lenses analyzed
spectroscopically in \citet{Momcheva2005}, four of which were targeted
there because they had known groups or good evidence for external shear.  
We use no information about environment to select the other
four lenses in the present sample.

\subsection {Data collection}

We obtained observations over several observing runs between 2002 and
2004 with the Mosaic-1 imager on the KPNO Mayall 4-m telescope and the
Mosaic II imager on the CTIO Blanco 4-m telescope.  The observing log
is given in Table \ref{tab.obs}.  The wide field of view of the Mosaic
cameras ($\sim 36\arcmin \times 36\arcmin$) covers the full virial
radius, $\rvir \approx 500 h^{-1} \, {\rm kpc}\, \sim 3\arcmin$, of any
groups that may be present around the lens galaxies, in addition to
all other massive structures that may affect the lens potential
\citep[see][]{Momcheva2005}.  We imaged each field with the ``nearly
Mould'' $I$ filter, and with either Harris
$V$ or Harris $R$, with the choice of bluer filter based upon the
location of the 4000\AA\ break at the lens galaxy redshift.  We
changed from $V$ to $R$ for $\zlens\geq 0.35$, approximately the
redshift at which the 4000\AA\ break enters the $R$ filter.\footnote{We have
since modified this strategy such that our ongoing survey now uses $V$
out to $\zlens= 0.5$, as the $V$ filter gives better redshift
resolution than the $R$ filter out to these larger redshifts.
However, in the present paper, the $R-I$ color proves sufficient to
identify groups for $0.35\leq \zlens\leq 0.5$, both local to the lens
and projected along the line-of-sight, as shown in
\S\ref{sec.algorithm.tests}.}

\begin{deluxetable*}{clllcl}
\tablewidth{0pt}
\tablecolumns{6}
\tablecaption{Observing Log\label{tab.obs}}
\tablehead{\colhead{Night} & \colhead{UT Date} & \colhead{Telescope} & 
  \colhead{Instrument} & \colhead{Seeing (\arcsec)} & \colhead{Comments}}
\startdata
K1 & 2002 March 14 & KPNO Mayall 4m & Mosaic-1 & 1.3 & \nodata \\
K2 & 2002 March 15 &                &          & 1.6 & \nodata \\
K3 & 2002 March 16 &                &          & 1.2 & \nodata \\
K4 & 2002 March 17 &                &          & 1.3 & cirrus; non-photometric \\
K8 & 2004 June 15 &                 &          & 1.0 & \nodata \\
C1 & 2002 May 10 & CTIO Blanco 4m & Mosaic II & $0.8-1$ & \nodata \\
C2 & 2002 May 11 &                &           & 1.0 & non-photometric \\
C3 & 2002 May 12 &                &           & 1.0-1.8 &
non-photometric, volatile seeing\\
C5 & 2003 December 14 &           &           & 0.8-1.0 & \nodata \\
C7 & 2003 December 18 &           &           & 0.8-1.0 & \nodata \\
C8 & 2003 December 20 &           &           & 0.8-1.0 & \nodata \\
C9 & 2004 June 18 &               &           & 0.9-1.2 & \nodata \\
\enddata
\end{deluxetable*}

In order to fill the gaps between the CCDs in the Mosaic cameras, we
dithered the exposures. During the 2002 runs, no specific dither
pattern was used; in subsequent runs we used a five-point dither
pattern.  We based our exposure times in 2002 ($\sim 1800$ sec in $R$ and
$I$) on those necessary to obtain S/N$=5$ at $I\approx 21.5$.
Subsequently, we found this signal-to-noise to be insufficient for
star-galaxy separation for $I\gtrsim 21$, and thus we increased the
total exposure times to 2400--2700 secs.  We split the 2002 exposures into
two 900 sec exposures in each band.  This strategy proved less than
ideal, as the signal-to-noise in the areas affected by the chip gaps
was appreciably lower, some small gaps in areal coverage remained, and
the $I$-band sky brightness was sufficiently high ($\sim$11000 ADU) to
fill the pixel wells to $\sim$30\% capacity, appreciably reducing
dynamic range.  In subsequent observing runs, we split the total
exposure time into 480 sec or 540 sec exposures.  This improved strategy
increased the signal-to-noise in the chip gap regions and kept the sky
levels at reasonable values.

\subsection{Data Reduction} 

We reduced the data using the \emph{MSCRED} package of
IRAF.\footnote{IRAF is distributed by the National Optical Astronomy
Observatories, which are operated by the Association of Universities
for Research in Astronomy, Inc., under cooperative agreement with the
National Science Foundation.}  Our data reduction closely followed the
prescription used by the NOAO Deep Wide-Field Survey as outlined in
the online notes by Jannuzi, Claver, \&
Valdes.\footnote{\url{http://www.noao.edu/noao/noaodeep/ReductionOpt/frames.html}}
In short, we combined each night's calibrations (bias frames, dome
flats, and night-sky flats) and applied these to the science images.
For KPNO data, we also created a map of the ``pupil-ghost'' image and
subtracted this from each frame.  For each night's $I$-band data, we
created, scaled, and subtracted fringing maps from each frame.  At
this point, we used coordinates from the Guide Star Catalog 2 to
refine the image world coordinate systems.\footnote{The Guide Star
Catalog was produced at the Space Telescope Science Institute under
U.S. Government grant. These data are based on photographic data
obtained using the Oschin Schmidt Telescope on Palomar Mountain and
the UK Schmidt Telescope.} The images were projected onto the tangent
plane and then stacked using $3\sigma$ clipping.  Unlike the Deep
Wide-Field Survey reduction routine, we did not remove cosmic rays in
a separate step, as the $\sigma$ clipping removed the vast majority of
hits.  We also did not fit and remove sky gradients prior to the image
stacking, as the resulting stacked images were generally cosmetically
better without the sky removal.  We set pixels in the final stacked
image that had no good values (i.e., unobserved regions, saturated
pixels, and bleed trails) to an arbitrary, high value for ease of
flagging in the analysis.

\emph{Photometric measurements} --- We obtained the lens field
photometry using SExtractor v.2.3.2 \citep{Bertin1996}.  Using the
image in the best-seeing band (usually $I$) as a reference, we
shifted and scaled the stacked images for a given field with the IRAF
\emph{geotran} task.  We convolved the best-seeing image with a
Gaussian kernel to degrade the seeing to that of the other band.

Total magnitudes are \citet{Kron1980} magnitudes given by the
SExtractor MAG\_AUTO  measured from the unconvolved
$I$-band image.  We measured galaxy colors via aperture magnitudes in
the matched-seeing images. We selected the aperture size to be a $6.1$
kpc physical radius, equivalent to 5 pixels ($=1\farcs3$) for a
$z=0.5$ galaxy. This aperture corresponds to the full width-half
maximum (FWHM) of the point-spread function (PSF) in the worst-seeing
lens field in this sample.

\emph{Star-galaxy separation} --- We tested three methods for
star-galaxy separation: (1) the stellarity index output by SExtractor,
which compares observed profiles with an input FWHM value using a
neural network algorithm, (2) $\chi^2$ measurements output by DAOphot
\citep{Stetson1987}, which are based on the goodness-of-fit of a model
PSF, and (3) the difference in aperture magnitudes between two concentric
apertures, with radii approximately equal to the FWHM and twice the
FWHM.  The last of these methods is motivated by the fact that point sources
have a universal PSF and thus should have a constant difference in
aperture magnitudes, while extended objects (galaxies) should not. We
inserted artificial stars and galaxies into the PG1115+080 field using
the IRAF \emph{artdata} routine, recovered these objects with
SExtractor, and performed star-galaxy separation using each of the
three methods.

The DAOphot analysis performed worst at star-galaxy separation due to
the known high-frequency variations in the Mosaic PSF.\footnote{See
http://www.physics.nau.edu/\~{}pmassey/Mosaicphot.html}
The other two methods worked comparably well over a wide range of
parameters, but the aperture method was able to go slightly fainter.

The SExtractor stellarity index is a continuous parameter running from
1 for point sources to 0 for obviously extended sources, with
intermediate values indicating various degrees of ambiguity.  This
parameter has two main drawbacks. First, it is quite sensitive to the
input value of the FWHM for faint objects.  Second, the value of the
stellarity index corresponding to the dividing line between stars and
galaxies is ill-defined.  The division between stars and galaxies is
subjective, and many different cuts exist in the literature.

\begin{figure}
\plotone{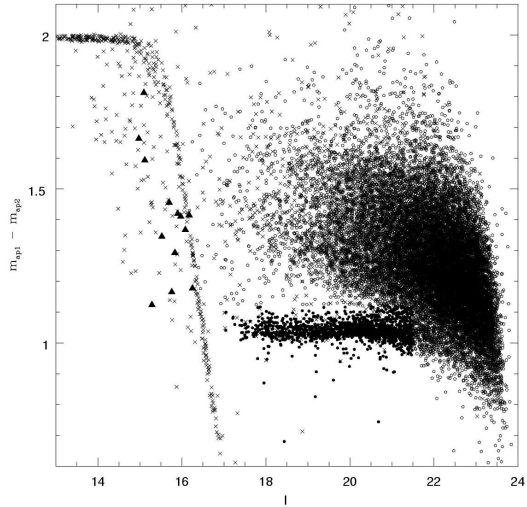}
\caption{Difference in aperture magnitudes for two concentric
  apertures as a function of magnitude in the field of B1422.  Crosses
  indicate objects that are flagged as saturated or have otherwise bad
  photometry; filled circles are objects identified as stars; open
  symbols are galaxies; and filled triangles are objects that we then
  classified by eye.  A visual inspection of the saturated objects
  finds only one bright galaxy near the edge of the field; the other
  objects are either saturated or bad measurements (e.g. near the
  image edge, under a bleed trail, a defect, etc.).  The star-galaxy
  separation routine works well for $16.5\lesssim I\lesssim
  21$. \label{fig.stargalsep}}
\end{figure}

The concentric-aperture method is illustrated in
Figure~\ref{fig.stargalsep}.  This method was robust over a large
magnitude range, though it failed at bright magnitudes due to an
ambiguity between bright extended objects and saturated stars.
Although SExtractor output saturation flags, some saturated stars
failed to be flagged.  Therefore, we classified by eye all non-flagged
objects brighter than the magnitude at which saturation set in. For
most fields, this involved fewer than a dozen objects.  Due to its
superior ability to identify unresolved objects, we therefore used the
concentric-aperture method for star-galaxy separation.

\subsection{Calibrations}

We calibrated our data obtained on photometric nights to the Kron-Cousins
filter set using Mosaic
images of Landolt standard fields \citep{Landolt1992}. We performed
multiple-aperture photometry of photometric standards using DAOphot.
Aperture corrections to infinity were calculated using the profile
fitting routine DAOgrow.  We assumed color transformations of
the form:
\begin{eqnarray}
v & = & V + a_0 + a_1 (V-I) + a_2 (X-1.25) \label{eqn.v}\\
r & = & R + b_0 + b_1 (R-I) + b_2 (X-1.25) \label{eqn.r}\\
i & = & I + c_0 + c_1 (V-I) + c_2 (X-1.25) \label{eqn.i}\, ,
\end{eqnarray}
where $v$, $r$, and $i$ are the instrumental magnitudes and $X$ is the
airmass.  In a few cases, depending on the bands observed during a
night, the color term used in Eqn.~\ref{eqn.v} was \bv, and the color
term used in Eqn.~\ref{eqn.i} was $R-I$; these are noted in
Table~\ref{tab.photcoeff}.  We assumed extinction values of
$a_2=0.15$, $b_2=0.10$, and $c_2=0.07$ magnitudes per unit airmass for
the Kitt Peak data \citep{Massey2002}, as we lacked sufficient
standard star observations to determine these values.  We were able to
calculate extinction coefficients for each night of CTIO data.  We
determined color terms $a_1$, $b_1$ and $c_1$ for each telescope run
and made no attempt to account for known, small (millimag)
chip-to-chip variations in color terms.  The zero points $a_0$, $b_0$
and $c_0$ were determined individually for each night.  The adopted
photometric coefficients are given in Table~\ref{tab.photcoeff}.

\begin{deluxetable*}{lccccc}
\tablewidth{0pt}
\tablecolumns{6}
\tabletypesize{\footnotesize}
\tablecaption{Adopted Photometric Coefficients\label{tab.photcoeff}}
\tablehead{
\colhead{Night\tablenotemark{a}} & \colhead{Band} & \multicolumn{3}{c}{Fit
  Coefficients\tablenotemark{b}} & \colhead{$N$}}
\startdata
K1 & $V$ & $a_0=-25.163\pm 0.002$ & $a_1=\phantom{-}0.034\pm 0.002$ & $a_2=0.15\tablenotemark{c}$ &  4 \\
   & $R$ & $b_0=-25.369\pm 0.005$ & $b_1=-0.052\pm 0.006$ & $b_2=0.10\tablenotemark{c}$ &  4 \\
   & $I$ & $c_0=-24.774\pm 0.010$ & $c_1=-0.013\pm 0.004$ & $c_2=0.07\tablenotemark{c}$ &  3 \\
K2 & $V$ & $a_0=-25.101\pm 0.007$ & $a_1=\phantom{-}0.034\pm 0.002$ & $a_2=0.15\tablenotemark{c}$ & 15 \\ 
   & $R$ & $b_0=-25.390\pm 0.006$ & $b_1=-0.052\pm 0.006$ & $b_2=0.10\tablenotemark{c}$ & 10 \\
   & $I$ & $c_0=-24.777\pm 0.003$ & $c_1=-0.013\pm 0.004$ & $c_2=0.07\tablenotemark{c}$ & 11 \\
K3 & $V$ & $a_0=-25.142\pm 0.004$ & $a_1=\phantom{-}0.034\pm 0.002$ & $a_2=0.15\tablenotemark{c}$ &  4 \\ 
   & $R$ & $b_0=-25.343\pm 0.003$ & $b_1=-0.052\pm 0.006$ & $b_2=0.10\tablenotemark{c}$ &  6 \\
   & $I$ & $c_0=-24.777\pm 0.003$ & $c_1=-0.013\pm 0.004$ & $b_2=0.07\tablenotemark{c}$ & 11 \\
K8 & $V$ & $a_0=-25.077\pm 0.023$ & $a_1=\phantom{-}0.001\pm 0.023$ & $a_2=0.15\tablenotemark{c}$ & 18 \\
   & $R$ & $b_0=-25.303\pm 0.021$ & $b_1=-0.081\pm 0.040$ & $b_2=0.10\tablenotemark{c}$ & 13 \\
   & $I$ & $c_0=-24.732\pm 0.017$ & $c_1=-0.031\pm 0.016$ & $c_2=0.07\tablenotemark{c}$ & 14 \\
C1 & $R$ & $b_0=-25.667\pm 0.011$ & $b_1=\phantom{-}0.013\pm 0.027$ & $b_2=0.031\pm 0.045$ &  8 \\
   & $I$ & $c_0=-24.912\pm 0.005$ & $c_1=\phantom{-}0.027\pm 0.012$\tablenotemark{d} & $c_2=0.000\pm 0.022$ & 10 \\
C5 & $V$ & $a_0=-25.330\pm 0.003$ & $a_1=\phantom{-}0.054\pm 0.005$ & $a_2=0.145\pm 0.006$ & 72 \\
   & $R$ & $b_0=-25.517\pm 0.003$ & $b_1=-0.022\pm 0.006$ & $b_2=0.097\pm 0.006$ & 77 \\
   & $I$ & $c_0=-24.860\pm 0.004$ & $c_1=\phantom{-}0.028\pm 0.004$ & $c_2=0.076\pm 0.007$ & 73 \\
C7 & $V$ & $a_0=-25.307\pm 0.004$ & $a_1=\phantom{-}0.036\pm 0.003$\tablenotemark{e} & $a_2=0.068\pm 0.036$ & 58 \\
   & $R$ & $b_0=-25.489\pm 0.004$ & $b_1=-0.010\pm 0.005$ & $b_2=0.087\pm 0.049$ & 26 \\
   & $I$ & $c_0=-24.829\pm 0.006$ & $c_1=\phantom{-}0.027\pm 0.004$ & $c_2=0.065\pm 0.072$ & 46 \\
C8 & $V$ & $a_0=-25.341\pm 0.008$ & $a_1=\phantom{-}0.043\pm 0.007$\tablenotemark{e} & $a_2=0.135\pm 0.024$ & 39 \\
   & $R$ & $b_0=-25.521\pm 0.006$ & $b_1=-0.029\pm 0.010$ & $b_2=0.092\pm 0.015$ & 32 \\
   & $I$ & $c_0=-24.872\pm 0.006$ & $c_1=\phantom{-}0.028\pm 0.005$ & $c_2=0.046\pm 0.017$ & 53 \\
C9 & $R$ & $b_0=-25.610\pm 0.006$ & $b_1=-0.015\pm 0.013$ & $b_2=0.090\tablenotemark{c}$ & 45 \\
   & $I$ & $c_0=-24.963\pm 0.007$ & $c_1=\phantom{-}0.070\pm 0.015$\tablenotemark{d} & $c_2=0.050\tablenotemark{c}$ & 23 \\
\enddata
\tablenotetext{a}{Photometric nights only}
\tablenotetext{b}{See Eqns.~\ref{eqn.v} to \ref{eqn.i}}
\tablenotetext{c}{Adopted from published mean extinction}
\tablenotetext{d}{Coefficient for $R-I$}
\tablenotetext{e}{Coefficient for \bv}
\end{deluxetable*}

We applied the zero points and extinction coefficients to calibrate
the galaxy photometry and colors but did not apply the color terms.
The color terms derived above are for stellar spectral energy
distributions (SEDs), which are roughly blackbody SEDs spanning
a range of temperatures.  The application of these color terms to
colors of redshifted stellar systems was not appropriate, and at
present we do not have calibrated spectrophotometry of any of our
targets to determine empirical color terms for these galaxies.  If we
assume that these color terms are similar in size to those in Table
\ref{tab.photcoeff}, then the systematic errors we introduced into the
colors would be on the order of 0.1 to 0.2 mags, similar to the size
of the random errors (\S\ref{sec.complete}).  For galaxies with
similar spectral energy distributions at a similar redshift, such as
those in the red sequence (\S\ref{sec.algorithm}), the systematic
offset is the same for each galaxy and will not affect the detection of the
red sequence.  Therefore, we simply acknowledge that no color term correction
has been applied to the photometry, likely introducing some small
systematics into the derived colors ($\lesssim 0.1$ mags) 
and into the photometry-based group redshifts.

In the cases where fields were imaged on non-photometric nights, we
obtained short calibration exposures of the fields on later
photometric nights.  We derived local standard stars in each field
using aperture photometry, including appropriate aperture corrections.
From these local standards we determined zero points, which include
the atmospheric extinction implicitly, for each field.  The fields
calibrated in this manner, and the photometric nights used for
calibration, are indicated in Table~\ref{tab.sample}.

Special calibration steps were required for the $V$-band imaging of
PG1115, for which we failed to obtain an image during a photometric
night. Our attempts to calibrate the field with archival \emph{HST}
and archival \emph{CFHT} data failed due to a lack of common
unsaturated stars between the archival data and our field.  Luckily,
the PG1115 field was included in the Third Data Release of the Sloan
Digital Sky Survey \citep[SDSS
DR3,][]{Abazajian2005}.\footnote{Funding for the Sloan Digital Sky
Survey (SDSS) has been provided by the Alfred P. Sloan Foundation, the
Participating Institutions, the National Aeronautics and Space
Administration, the National Science Foundation, the U.S. Department
of Energy, the Japanese Monbukagakusho, and the Max Planck
Society. The SDSS Web site is \url{http://www.sdss.org/}.}  We
retrieved PSF magnitudes for all stellar objects within a
15\arcmin~radius of PG1115 from the SDSS DR3 database.  We transformed
these magnitudes to $U\!BV\!R_CI_C$ using the observed transformations
given in \citet{Smith2002}; we then used these magnitudes as local
standards in our $V$ and $I$ images.  The resulting zero-points were
$a_0 = 25.390\pm 0.018$ for the $V$-band and $c_0=24.947\pm0.052$ for
the $I$-band.  The $I$-band zero point was consistent with that
derived from the Landolt standards, 
giving us confidence that the $V$-band calibration was accurate.

\subsection{Completeness and Photometric Accuracy\label{sec.complete}}

We determined the completeness and photometric accuracy of the
observations via artificial galaxy tests.  Artificial galaxies were
placed at random locations in the field, with additional artificial
galaxies placed at random within $\rvir\equiv 500 h^{-1}\,{\rm kpc}$
of the lens galaxy. The input
galaxy colors were chosen to match those expected for an early-type
galaxy at the lens redshift (see Appendix A). The
input magnitudes were distributed randomly and equally between $I_*-2$
and $I=22.5$, where $I_*$ is the observed apparent $I$ magnitude of an
$L_*$ galaxy located at \zlens (see Appendix A).  For
the lenses in this paper, $17.40 \leq I_* \leq 19.26$.  This range
includes the bulk of the galaxies in our photometric catalog.

Galaxy luminosity profiles were de Vaucouleurs ($r^{1/4}$) profiles
convolved with a Gaussian kernel whose FWHM matched that measured in
the image. We calculated an effective radius from the Kormendy
relation
\begin{equation}
\avg{\mu}_e=\alpha + \beta\log R_e\, ,
\end{equation}
where $R_e$ is the effective radius in kpc and $\avg{\mu}_e$ is the
average surface brightness interior to $R_e$ in units of magnitudes
per square arcsecond \citep{Kormendy1977}.  By definition, the total
light within $R_e$ is half of the total galaxy luminosity, so
\begin{equation}
\avg{\mu}_e = 2.5\log(2\pi) + 5\log(r_e) + m_T\, ,
\end{equation}
where $r_e$ is the effective radius in units of arcseconds and $m_T$
is the total apparent magnitude of the galaxy.  Combining these two
equations, setting $\beta=3$ \citep[the approximate value from
][]{Barbera2003}, and solving for $\log R_e$, we get
\begin{equation}
\log R_e = \frac{\alpha}{2} - 1.25\log(2\pi) - \frac{5}{2}\log
\theta_{\rm kpc}- \frac{m_T}{2}\, , \label{eqn.kormendy}
\end{equation}
where $\theta_{\rm kpc}$ is the angle (in arcseconds) subtended by 1
kpc at the lens galaxy redshift.  For $\alpha$, \citet{Barbera2003}
give a rest-frame $V$-band value of
\begin{equation}
\alpha_V = 19.05 + 7.7\log (1+\zlens) + A_V\, ,\label{eqn.kormenalpha}
\end{equation}
where $A_V$ is the $V$-band extinction.  We converted this value to
the observed $I$-band with a $K$-correction computed with the
\emph{synphot} tools in IRAF's \emph{stsdas} package, using the E/S0
spectrum from \citet{Coleman1980}.  We also applied Galactic reddening
corrections interpolated from the \citet{Schlegel1998} maps using the
reddening curve of \citet{Rieke1985}.

We inserted the resulting artificial galaxies into each lens field and
recovered them using SExtractor.  Each run included 365 artificial
galaxies: 350 randomly scattered throughout the field and an
additional 15 within \rvir of the lens.  We considered an artificial
galaxy recovered if its centroid was within 2.5 pixels and its $I$
magnitude and color were within 1.0 mag of the input values.  Fifty
runs were performed in each lens field to build up statistics.

\begin{figure*}
\plotone{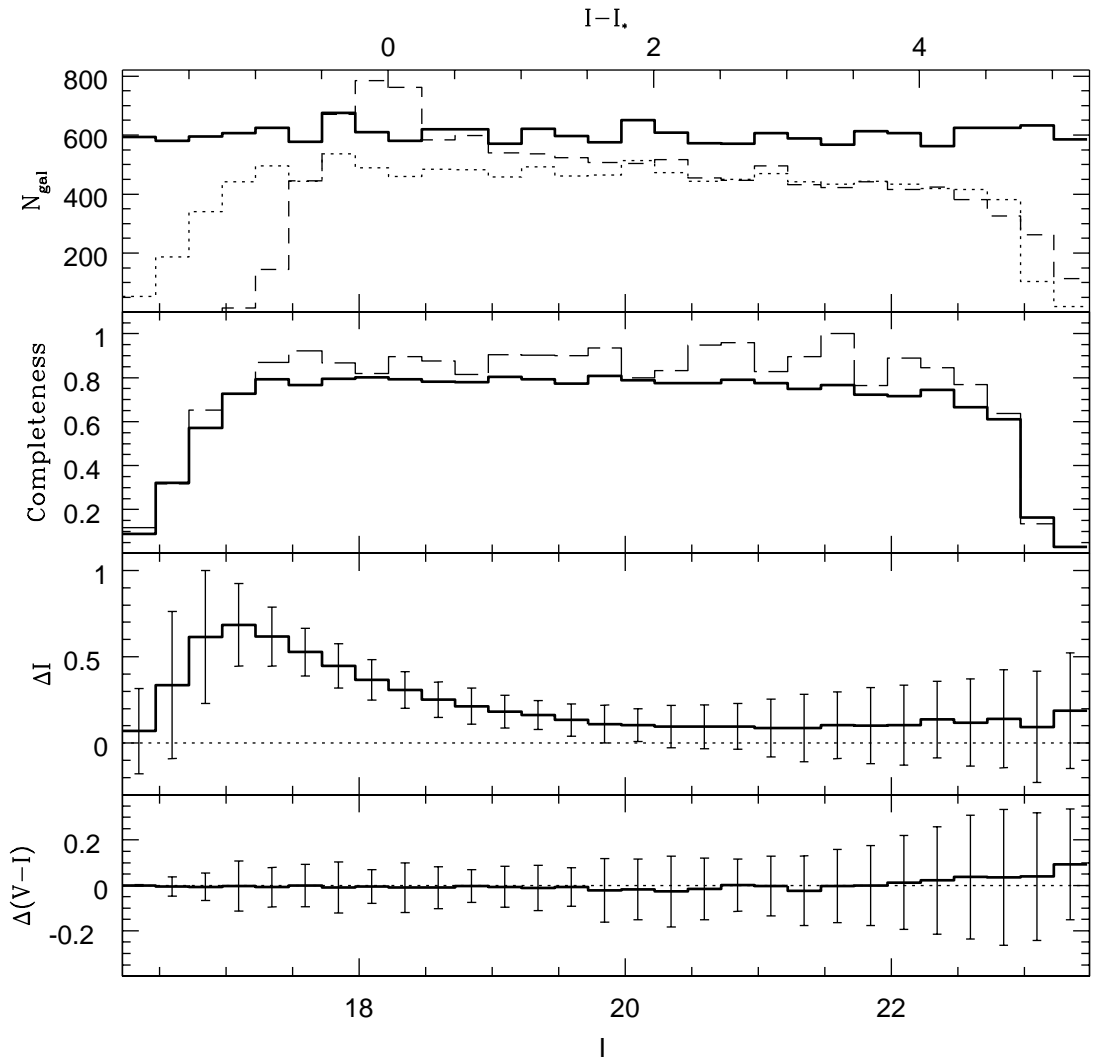}
\caption{Results of artificial galaxy tests in the field of CLASS
  B1422+231 (hereafter B1422).  The top panel shows the number of
  artificial galaxies (over all 50 runs) as a function of input total
  magnitude for all galaxies (heavy solid line) and all recovered
  galaxies (dotted line), as well as the number of recovered galaxies
  as a function of \emph{measured} magnitude (dashed line).  The
  second panel shows the completeness as a function of input total
  magnitude for the entire image (heavy solid line) and the region
  within a $500h^{-1}$ kpc projected radius of the lens (dashed line).
  The third panel shows the mean difference between the total
  magnitude and the measured magnitude as a function of total
  magnitude; error bars give the $1\sigma$ deviation about the mean.
  The bottom panel shows the mean difference between the input and
  measured aperture colors as a function of total magnitude; error
  bars are the $1\sigma$ deviation about the mean.\label{fig.artgal}}
\end{figure*}

We plot the results from the artificial galaxy tests in a typical
field (B1422) in Figure \ref{fig.artgal}.  Two systematic issues are
readily observed.  The recovered galaxy magnitudes consistently
underestimated the flux, the effect being greatest for the brightest
galaxies.  This underestimate is due to the low surface brightness
outer regions of an elliptical galaxy not being detected by SExtractor.  The
effect is strongest at bright luminosities because intrinsically
bright ellipticals have the most diffuse surface brightness profiles.
The net effect is to amplify the number of galaxies around $I_*$ and
suppress the number of galaxies brighter than $\sim I_*-1$.  For
galaxies in these magnitude ranges, another method of measuring
magnitudes (such as profile fitting) is necessary to accurately
determine the total luminosity.  Nevertheless, we did not attempt to
correct the photometric catalog for these systematic magnitude offsets
for two reasons.  First, our method of measuring magnitudes is in
common use; correcting to total magnitudes would hinder comparisons
with other samples.  Second, the exact offset depends on the input
galaxy luminosity profile.  Our tests were limited to bulge-dominated
galaxies, so we could not estimate the corrections for disk-dominated
galaxies.  In future work we will perform detailed profile fits to the
galaxies near the lens systems to address this systematic effect
carefully.

For the brightest bins in the figure, the completeness is low because
the central regions of bright galaxies are flagged as saturated. In
order to determine if this saturation affects our results, we visually
inspect objects flagged as saturated in a subset of the fields.  While
a handful of bright galaxies in each field are flagged as saturated,
in no case are the flagged galaxies potential group members; these
galaxies are either more than several arcminutes from the lens or
obviously foreground, with magnitudes significantly brighter and
physical extents significantly larger than any other galaxy near the
lens.

Given that galaxies in groups tend to be clustered, we test whether
crowding (i.e., the blending of overlapping galaxies) could be a
significant source of incompleteness in our data.  Overall, we find
that the completeness within \rvir of the lens is similar to that in
the remainder of the field.  In some fields, such as B1422 (Figure
\ref{fig.artgal}), we find the completeness within \rvir of the lens
is slightly higher than in the field as a whole. In other fields the
completeness is slightly lower.  These small completeness variations
are most likely due to statistical fluctuations resulting from the
smaller number of artificial galaxies used to calculate the
completeness in the relatively small area surrounding the lens as
compared to the much larger area of the complete Mosaic field. We
therefore conclude that crowding is not an issue in our data.

Given the uncertainties in the model galaxies, the (ignored) scatter
in the Kormendy relation, and the fact that the input artificial
galaxies are solely early-type galaxies at a fixed redshift, the
completeness is not necessarily valid outside the red sequence and is
subject to uncertain systematics.  However, the observed
field-to-field stability of the completeness, despite differing lens
redshifts, seeing, and field depth, suggests that our completeness
calculations are relatively robust.

\section{Identifying Groups and Candidate Group Members\label{sec.findgroups}}

Due to the morphology-density relation, the early-type fraction is
higher in overdense regions like groups and clusters of galaxies than
elsewhere.  In such cases, one observes a ``red sequence'' (RS) of
elliptical and S0 galaxies in the color-magnitude diagram (CMD) of
galaxies in the region.  The RS forms the basis for our
successful two-band photometric detection of galaxy groups and
clusters.

Our RS detection algorithm is a modified version of that presented by
\citet{Gladders2000}.  Their algorithm searches color and astrometric
space for localized enhancements of early-type galaxies at similar
redshifts and has proven successful in locating group-mass
structures over a wide range in redshift using data from optical
imaging surveys.  As our survey is targeted toward the immediate
environment of lenses, we do not need to include the spatial filtering
in our selection.  The other major difference between our algorithm
and that of \citet{Gladders2000} is that we include human interaction
in the selection process.  While this introduces a hard-to-quantify
bias into the resulting group sample, it enables us to detect
less-massive groups ($\sigma\lesssim 300\,{\rm km\, s}^{-1}$).  As
shown in \S\ref{spec.comp}, the resulting group catalog is nearly
complete with few false detections, suggesting that any bias
introduced by the human interaction is small.

\subsection{Group Detection\label{sec.algorithm}}

Our basic RS detection algorithm is as follows.  First we search
within \rvir of the lens galaxy for an excess of galaxies with similar
colors above the measured background color distribution. We fit an RS
to these galaxies, and calculate the projected spatial centroid of the
candidate group.  Using the RS fit and centroid, we repeat this search
to ensure that the group is not spurious and to refine the RS
parameters and centroid measurement.  We continue the iterative
process until the changes in the centroid and the RS parameters are
within the calculated errors.  The first time we apply this algorithm
to a field, we focus on galaxy colors near that of an early-type
galaxy at the lens redshift, \zlens.  We then search for line-of-sight
groups by repeating the algorithm until we have investigated all peaks
in the initial color distribution.

We assume an RS of the form 
\begin{equation}
\label{eqn.rs}
(V-I)_0 = (V-I)_* - \beta_V(I_0-I_*)\, ,
\end{equation}
where $\beta_V$ is the slope of the RS, and $(V-I)_*$ is the
unreddened observed $V-I$ color of an $L_*$ galaxy at the lens
redshift.  For fields with $R$-band imaging, we use this same
relation with $V$ replaced by $R$.

We allow both the RS slope $\beta_V$ and the group redshift, which
determines $I_*$ and $(V-I)_*$, to vary freely. We calculate the
initial $(V-I)_*$ and $I_*$ from the galaxy models described in
Appendix A and the known \zlens.  The RS slope has
been found to have values around $d(U-V)/dM_V = 0.08$
\citep[e.g.,][]{Bower1992,Bell2004,McIntosh2005}.  Because the
color-magnitude relation is difficult to model properly, and because
our bandpasses do not correspond exactly to rest-frame $U$ and $V$, it
does not necessarily follow that this slope can be directly compared
with the $V-I$ or $R-I$ slopes of the groups in this paper. However,
we show in \S\ref{sec.results.sample} that the actual fit slopes from
our sample are indistinguishable from this value.  We therefore
initially set $\beta_V = 0.08$.  Although this could bias the final
fit slopes, we show in \S\ref{sec.algorithm.tests} that we can recover
the correct RS slope even if this initial assumed value is incorrect.

For the first iteration of our RS search, we limit the search to
within \rvir($=500 h^{-1}\,{\rm kpc}$) of the lens galaxy.  This radius
is large enough ($\sim\!2\arcmin$--$3\arcmin$) to include most of the
line-of-sight structures that could impact lens models
\citep[see][]{Momcheva2005}.  It is possible that a massive galaxy
cluster could lie outside this radius and still alter the lens
potential, so we estimate the fraction of lenses affected by such a
cluster as follows.  We consider a lens model perturbation significant
if the convergence $\kappa\geq 0.05$.  From Eqn.~A20 in
\citet{Momcheva2005}, we find that the maximum impact parameter
$b_{\rm max}$ that a cluster can have and significantly impact the
lens model is $b_{\rm max} = 4.8\times 10^{-6} \sigma^2$, where
$b_{\rm max}$ is in arcminutes and $\sigma$ is the cluster velocity
dispersion in km s$^{-1}$.  This conservatively assumes that the
cluster is at the lens redshift; for clusters at other redshifts,
$b_{\rm max}$ will be smaller.  For a galaxy cluster with $\sigma\sim
750$ km s$^{-1}$, the median velocity dispersion for rich Abell
clusters \citep{Zabludoff1990}, $b_{\rm max} \approx 3\arcmin$ ---
close to the radius within which we search anyway.  The surface
density of galaxy clusters is $\sim 11$ per sq. degree
\citep[][$0.3\lesssim z_{\rm cluster}\lesssim 0.9$, correcting for
their 30\% false detection rate]{Gonzalez2001}.  On average, this
results in 0.07 galaxy clusters within $b_{\rm max}$.  We consider
this to be an upper limit, given that we are already searching for
groups and clusters out to a significant fraction of $b_{\rm max}$ and
that $b_{\rm max}$ is smaller if there is any redshift difference
between the cluster and the lens.  We therefore conclude it is
unlikely that any significant perturbing structures lie outside our
search radius.

We correct galaxy magnitudes for interstellar extinction and reddening
using standard Galactic extinction \citep{Rieke1985} with $R_V=3.1$
and extinctions from \citet{Schlegel1998} obtained from the NASA
Extragalactic Database.\footnote{This research has made use of the
NASA/IPAC Extragalactic Database (NED) which is operated by the Jet
Propulsion Laboratory, California Institute of Technology, under
contract with the National Aeronautics and Space Administration.}
We do not attempt any corrections for differential Galactic extinction
across the field; any variations within \rvir of the lens are likely
to be negligible at the Galactic latitudes of these lenses. As is
common practice, extinction-corrected magnitudes and colors are
denoted by the subscript ``0.''

To account for the slope of the RS, we define an ``effective color''
for each galaxy:
\begin{equation}
(V-I)_{\rm eff} = (V-I)_0 + \beta_V(I_0-I_*)\, .
\end{equation}
This represents the color each galaxy would have if it resided on the
RS and had an apparent magnitude $I=I_*$.  This definition is useful
because all RS galaxies in a group should have the same effective
color.  Hence, we can identify RSs as peaks in a histogram of
effective color, providing an efficient way of searching for groups of
early-type galaxies.  The search for peaks in the effective color
histograms is also adept at identifying groups with high late-type
fractions, as blue galaxies in a group will all have roughly similar
colors, although in the absence of morphological or spectroscopic data
to classify the galaxies as late-types, the derived redshift of the group
will be incorrect (see \S\ref{sec.algorithm.tests}).

We compute effective color histograms for galaxies within \rvir of the
lens galaxy, as well as for all galaxies in the image (normalized to
the area within \rvir).  The number density of field galaxies as a
function of apparent magnitude has a steeper slope than typical group
luminosity functions \citep[e.g.,][]{Lin1999,Gladders2000}, so it is
possible to choose a magnitude limit that maximizes the contrast
between the RS and the field galaxy distribution.  This magnitude
limit will vary from field to field depending on the richness of the
group and statistical fluctuations in the number of background
galaxies.  Our empirical testing finds that, in most cases, the best
contrast between the group and the field is achieved for a limiting
magnitude of $I\approx I_*+2.5$.  We therefore consider only galaxies
brighter than $I_*+2.5$ throughout this algorithm.

We visually examine the effective histograms for an excess of galaxies of a
similar effective colors relative to the normalized
field background, an indication of a possible group.  We select all galaxies
within a distance $\Delta(V-I)_{\rm eff}$ of the peak in the effective
color histogram as candidate RS galaxies.  We
take $\Delta(V-I)_{\rm eff}$ to be the larger of the photometric color
error and 0.05 mags, the typical RS scatter in the literature
\citep[e.g.,][]{Bower1992,McIntosh2005}.  The photometric color errors
are typically larger than the intrinsic scatter for $I\gtrsim 21$.

We use linear regression to fit a linear RS to the selected galaxies
brighter than $I_*+2.5$, thereby refining the RS slope $\beta_V$ and
intercept.  From this fit we determine the new values of $(V-I)_*$,
$I_*$, and the RS redshift $z_{\rm RS}$ based on the early-type galaxy
photometric models (Appendix A).  As discussed in
Appendix A, we note that changes in the assumed
normalization of $I_*$ results in systematic changes in $z_{\rm RS}$,
though for the range of published values of $L_*$, this bias is
$\Delta z_{\rm RS} \leq 0.02$. We also calculate
the raw and luminosity-weighted group centroids of the RS galaxies.

At this point, we can either accept or reject the new parameter values
and then choose whether another iteration should be performed.
Changes in redshift cause the projected angular size of \rvir to be
recomputed, and changes in the centroids adjust the area of sky in
which galaxies are examined.  The iterative process continues until
the change in the RS redshift between iterations is less than $\Delta
z_{\rm RS} =0.02$ and the shift in the centroid is less than
$1\sigma$.  If multiple peaks in the color distribution are observed,
we repeat the entire process on each potential RS.  For the purpose of
this paper, we define ``significant'' candidate RSs as those with at
least three selected galaxies above the averaged background persisting
for at least two iterations, criteria that produce the best results in
tests with mock groups (see \S\ref{sec.algorithm.tests}).

\begin{figure*}
\includegraphics[angle=270,scale=0.67]{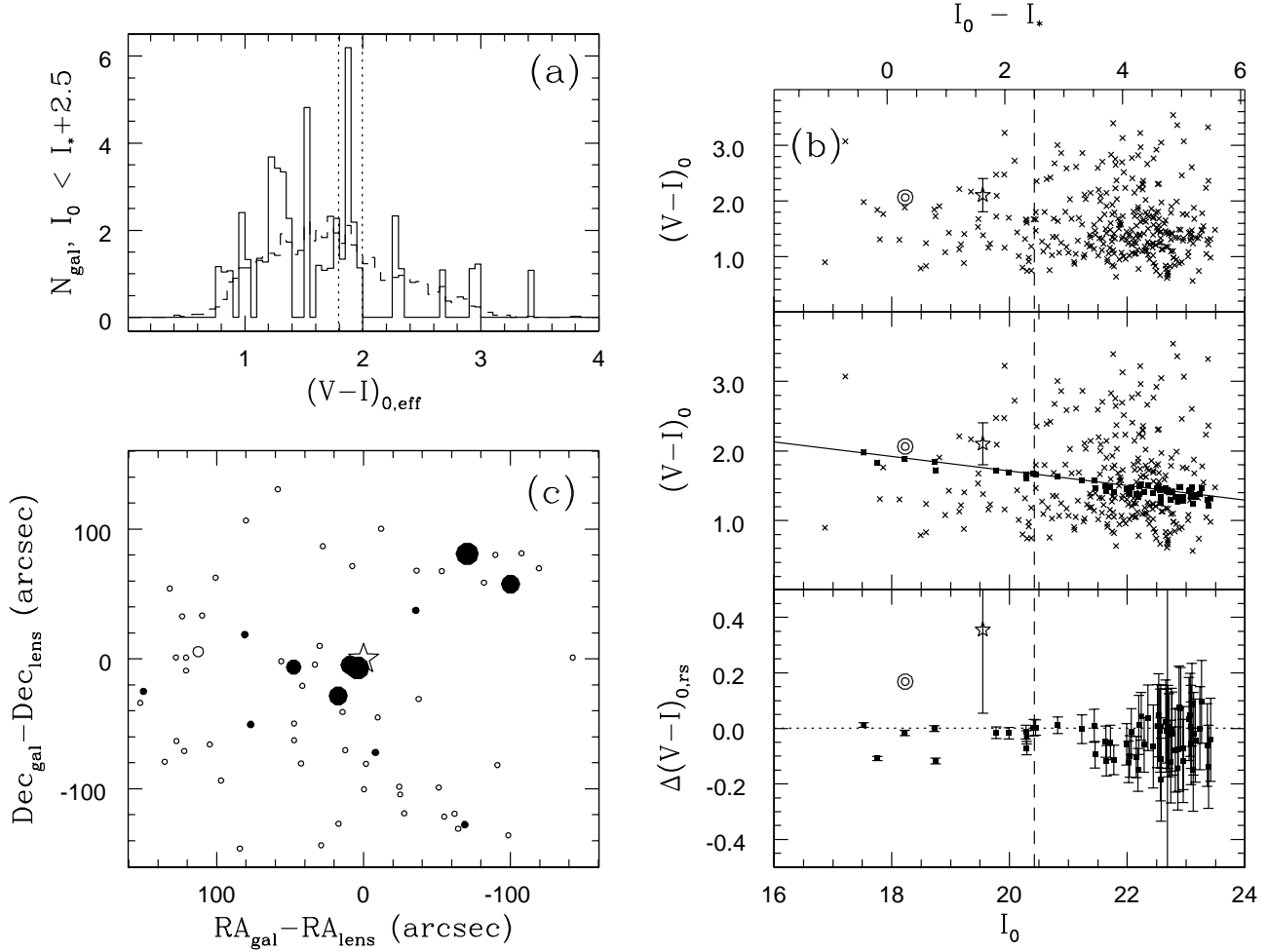}
\caption{Illustration of our group-finding algorithm for
  B1422. \emph{(a)} Histograms of effective colors for galaxies
  brighter than $I_*+2.5$ projected within \rvir (solid) and for the
  entire field normalized to the area within \rvir (dashed).  Vertical
  dotted lines bracket the region in $(V-I)_{\rm eff}$ from which the
  final group member catalog was selected. \emph{(b, top panel)}
  Color-magnitude diagram for all galaxies (crosses) projected within
  \rvir. The star with error bars indicates the lens galaxy;
  concentric circles indicate the expected location of an $L_*$
  early-type galaxy at the lens redshift. \emph{(b, middle panel)}
  Color-magnitude diagram for galaxies projected within \rvir
  (crosses) with galaxies selected by the red sequence finding
  algorithm shown as filled boxes.  The solid line is the fit red
  sequence. The vertical dashed line in all right-hand panels
  indicates the faint magnitude limit of the final red-sequence
  fitting. \emph{(b, bottom panel)} Residuals of the selected red
  sequence candidates about the best-fit line.  Error bars indicate
  errors estimated from the artificial galaxy tests.  
  \emph{(c)} Map of red sequence-selected
  galaxies projected within \rvir as a function of distance from the
  lens galaxy (open star). Filled circles indicate galaxies brighter
  than $I_*+2.5$ and are larger for brighter objects.  This figure illustrates
  the ability of the red sequence finding algorithm to locate physical groups.
\label{fig.b1422_ex}}
\end{figure*}

An example of this process applied to the lens B1422 is shown in
Figure~\ref{fig.b1422_ex}.  We see an RS of galaxies at $V-I\approx
1.9$, roughly the effective color of early-type galaxies at \zlens and
also the approximate color of the lens galaxy.  At brighter
magnitudes, the group luminosity function clearly dominates that of
the field.  A map of the galaxies in this range
(Fig.~\ref{fig.b1422_ex}c) shows that the collection of bright
galaxies, and therefore the group, is nearly centered on the lens
galaxy and is elongated.  This figure shows that our group detection
algorithm is capable of detecting physical groups and constraining the
group redshift, richness, centroid, and shape.  We note that some of
the bright galaxies in the selected RS may not be group members but
rather higher-redshift blue galaxies.  In the future, we will also
perform a morphological analysis, which should separate background
disk-dominated galaxies from the early-type RS group members. For now,
we make two tests of the algorithm, one using simulated groups of
galaxies (\S\ref{sec.algorithm.tests}), the other using spectroscopy
of eight of the twelve lens fields to verify RS membership
(\S\ref{spec.comp}).

\subsection{Testing the Algorithm: Simulated Data\label{sec.algorithm.tests}}  

We test our RS search algorithm by inserting mock galaxy groups into
each field's galaxy catalog.  We construct mock photometric 
groups from empirical luminosity functions.  The early-type fraction
and number of bright group galaxies are readily modified to simulate
galaxy groups of different richnesses. The precise luminosity function
does not significantly affect the success of the algorithm.

We begin by creating a mock group galaxy catalog designed to cover a
range of redshifts ($z=0.2,\,0.4,\, 0.6$), early-type fractions
($f_{\rm e}=0.0,\,0.5,\,1.0$), RS slopes ($\beta_V=$0.05, 0.10, 0.15),
and richnesses.  We define the mock group richness by the number of
galaxies brighter than $I_* + 2$, and our mock groups are designed to
have richnesses of 4, 8, or 16.  A richness of 4 corresponds to Local
Group analogues.  We derive cumulative luminosity functions from the
empirical group luminosity functions presented in
\citet{Zabludoff2000} to determine that richnesses of 8 and 16
correspond to velocity dispersions of $\sigma_{\rm v}\sim 300\,{\rm km
\,s}^{-1}$ and $\sigma_{\rm v}\sim 500\,{\rm km \,s}^{-1}$,
respectively, though there is significant scatter in this relation.

We select the luminosity function for each group from the empirical
luminosity functions derived from GEMS data \citep{Miles2004}.  That
study presents different luminosity functions for X-ray bright
($L_x\geq 10^{41.7}\,{\rm erg\,s}^{-1}$) and X-ray faint/undetected
groups.  This luminosity break corresponds to a velocity dispersion of
$\sim 200$ km s$^{-1}$\citep{Mulchaey1998}, so we use the GEMS
empirical X-ray faint luminosity function for groups with richness 4
and the X-ray bright luminosity function for groups with richness 8 or
16.

We assign each mock galaxy an $I$ magnitude drawn randomly from a
distribution defined by the appropriate luminosity
function.  We then label each galaxy an early- or late-type with a
probability based on the input early-type fraction ($f_{\rm
e}$). Late-type galaxies are further divided with equal probability
into Sa, Sb, and Sc types.  We assign colors based on the appropriate
photometric model (Appendix A) for the assigned morphology,
and, for early-types, we place the galaxy on the RS with a selected RS slope
between $\beta_V=0.05$ and 0.15.  We add Gaussian scatter of
$\sigma_{V-I}=0.05$ mag, similar to the observed scatter in other RS
studies, to each mock galaxy's color.  We also add additional scatter
to the $I$ magnitude and color of each galaxy to simulate the
photometric errors based on the photometric accuracy determined from
the artificial galaxy tests (\S\ref{sec.complete}).

We now need to assign positional data to each galaxy.  To simulate the
foreground and background galaxies in the field, we assign the
cataloged galaxies random $x$- and $y$-positions in the B1422 field
(chosen to be a representative lens field), thereby mixing any actual galaxy
groups and clusters into the field galaxy population.  We assume that
all mock group galaxies lie within \rvir of the lens, and we assign
these galaxies positions drawn randomly from a 2-dimensional Gaussian
density function centered on the lens position.  We then run the group
finding algorithm on the combined catalog.  We consider a mock group
to be recovered if we find a candidate RS with an effective color
within 0.2 of the input group's color, a value similar to our input
color errors.  This criterion prevents the detection of spurious
groups, i.e., those containing none of the input mock galaxies.  We
make an exception to this criterion if the group has a high late-type
fraction, in which case there often is a false RS created by having
many blue galaxies with similar colors.  In this circumstance, the
group is considered recovered despite having an effective color $>$0.2
mag bluer than that of a true RS at the group redshift.  In group
catalogs based on our actual data, groups with high late-type
fractions would be labeled line-of-sight groups.

Our knowledge of the mock group color could potentially bias the group
finding algorithm toward detection of the mock group.  However, we
also know the redshifts (and therefore expected RS colors) of the
lenses in the present sample, so any bias for detecting groups at
\zlens and the mock groups is the same.  The small number of
interloping groups identified from spectroscopy that are undetected
photometrically (see \S\ref{spec.comp}) argues that this bias is also
small for line-of-sight group detection.  The results of the
simulations are given in Table~\ref{tab.artgrps}.

\begin{deluxetable*}{ccccccccc}
\tablewidth{0pt}
\tablecolumns{9}
\tablecaption{Results from group-finding algorithm tests with mock groups\label{tab.artgrps}}
\tablehead{ \colhead{$z$} & \colhead{$f_{\rm e}$} & \colhead{$N_{I_*+2}$} & 
  \colhead{$\beta_V$} & \colhead{$N_{rs}$} & \colhead{$\overline{N}_{I_*+2,det}$} & 
  \colhead{$\overline{\Delta\beta_V}_{\rm fit}$}  & \colhead{$\overline{\Delta(V-I)}_{*,{\rm fit}}$} &
Scatter \\
\colhead{(1)} &\colhead{(2)} &\colhead{(3)} &\colhead{(4)} &\colhead{(5)} &\colhead{(6)} &\colhead{(7)} &\colhead{(8)} &\colhead{(9)} }
\startdata
0.2  & 1.00 & 16 & 0.05 & 5 & $13.3\pm  3.2$ & $ 0.023\pm 0.022$ & $-0.011\pm 0.015$ & $0.065\pm 0.019$ \\
     &      &    & 0.10 & 5 & $14.1\pm  2.4$ & $ 0.005\pm 0.014$ & $-0.027\pm 0.019$ & $0.080\pm 0.015$ \\
     &      &    & 0.15 & 5 & $14.4\pm  2.5$ & $-0.014\pm 0.011$ & $-0.022\pm 0.012$ & $0.067\pm 0.025$ \\
     &      &  8 & 0.05 & 4 & $ 6.5\pm  1.0$ & $ 0.017\pm 0.044$ & $\phantom{-}0.008\pm 0.027$ & $0.079\pm 0.016$ \\
     &      &    & 0.10 & 3 & $ 6.8\pm  3.2$ & $-0.038\pm 0.046$ & $\phantom{-}0.001\pm 0.029$ & $0.008\pm 0.002$ \\
     &      &    & 0.15 & 4 & $ 6.0\pm  1.6$ & $ 0.009\pm 0.043$ & $-0.007\pm 0.013$ & $0.005\pm 0.002$ \\
     &      &  4 & 0.05 & 1 & $10.6$ & $ 0.028$ & $ 0.030$ & $0.007$ \\
     &      &    & 0.10 & 3 & $ 3.9\pm  1.1$ & $-0.026\pm 0.032$ & $-0.036\pm 0.015$ & $0.097\pm 0.012$ \\
     &      &    & 0.15 & 0 & \nodata & \nodata & \nodata & \nodata \\
     & 0.00 & 16 &\nodata&5 & $10.7\pm  4.8$ & $ 0.062\pm 0.053$ & $-0.184\pm 0.013$ & $0.010\pm 0.005$ \\
     &      &  8 &\nodata&4 & $ 5.2\pm  1.0$ & $ 0.047\pm 0.034$ & $-0.179\pm 0.020$ & $0.013\pm 0.006$ \\
     &      &  4 &\nodata&3 & $ 3.2\pm  1.0$ & $ 0.104\pm 0.020$ & $-0.186\pm 0.008$ & $0.019\pm 0.004$ \\
     & 0.50 & 16 & 0.10 & 5 & $ 7.8\pm  2.9$ & $-0.001\pm 0.008$ & $-0.047\pm 0.024$ & $0.011\pm 0.004$ \\
     &      &  8 & 0.10 & 3 & $ 5.4\pm  0.9$ & $-0.029\pm 0.014$ & $-0.052\pm 0.043$ & $0.017\pm 0.004$ \\
     &      &  4 & 0.10 & 1 & $ 4.6$ & $ 0.012$ & $ 0.030$ & $0.011$ \\
0.4  & 1.00 & 16 & 0.05 & 5 & $15.2\pm  3.9$ & $ 0.026\pm 0.023$ & $-0.017\pm 0.016$ & $0.107\pm 0.017$ \\
     &      &    & 0.10 & 5 & $17.6\pm  3.1$ & $ 0.000\pm 0.010$ & $-0.011\pm 0.016$ & $0.087\pm 0.027$ \\
     &      &    & 0.15 & 5 & $16.2\pm  2.2$ & $-0.001\pm 0.027$ & $-0.023\pm 0.013$ & $0.088\pm 0.032$ \\
     &      &  8 & 0.10 & 5 & $ 7.9\pm  2.2$ & $ 0.017\pm 0.027$ & $-0.017\pm 0.016$ & $0.118\pm 0.026$ \\
     &      &  4 & 0.10 & 3 & $ 5.2\pm  3.1$ & $ 0.024\pm 0.054$ & $-0.031\pm 0.054$ & $0.146\pm 0.014$ \\
     & 0.50 & 16 & 0.10 & 5 & $ 8.4\pm  2.3$ & $ 0.034\pm 0.043$ & $-0.178\pm 0.219$ & $0.102\pm 0.026$ \\
     &      &  8 & 0.10 & 4 & $ 5.3\pm  1.8$ & $ 0.000\pm 0.067$ & $-0.244\pm 0.249$ & $0.125\pm 0.014$ \\
     &      &  4 & 0.10 & 1 & $ 3.1$ & $-0.041$ & $-0.538$ & $0.105$ \\
     & 0.00 & 16 & \nodata & 5 & $10.1\pm  4.0$ & $ 0.121\pm 0.048$ & $-0.441\pm 0.074$ & $0.107\pm 0.037$ \\
     &      &  8 & \nodata & 4 & $ 7.6\pm  2.8$ & $-0.013\pm 0.028$ & $-0.527\pm 0.057$ & $0.108\pm 0.008$ \\
     &      &  4 & \nodata & 0 & \nodata & \nodata & \nodata & \nodata \\
0.6  & 1.00 & 16 & 0.05 & 5 & $12.0\pm  3.4$ & $ 0.014\pm 0.016$ & $-0.009\pm 0.008$ & $0.109\pm 0.027$ \\
     &      &    & 0.10 & 5 & $12.0\pm  3.6$ & $ 0.007\pm 0.035$ & $-0.009\pm 0.016$ & $0.101\pm 0.026$ \\
     &      &    & 0.15 & 5 & $10.3\pm  1.7$ & $ 0.002\pm 0.036$ & $-0.001\pm 0.020$ & $0.107\pm 0.035$ \\
     &      &  8 & 0.10 & 1 & $ 4.8$ & $ 0.002$ & $-0.029$ & $0.097$ \\
     &      &  4 & 0.10 & 0 & \nodata & \nodata & \nodata & \nodata \\
     & 0.50 & 16 & 0.10 & 3 & $ 9.8\pm  5.7$ & $ 0.009\pm 0.037$ & $-0.010\pm 0.022$ & $0.130\pm 0.009$ \\
     &      &  8 & 0.10 & 2 & $ 5.1\pm  0.9$ & $-0.019\pm 0.018$ & $-0.336\pm 0.474$ & $0.124\pm 0.001$ \\
     &      &  4 & 0.10 & 2 & $ 3.4\pm  0.1$ & $ 0.003\pm 0.054$ & $-0.650\pm 0.030$ & $0.108\pm 0.013$ \\
     & 0.00 & 16 &\nodata& 4 & $ 8.8\pm  4.0$ & $ 0.044\pm 0.088$ & $-0.507\pm 0.088$ & $0.109\pm 0.020$ \\
     &      &  8 &\nodata& 3 & $ 4.1\pm  0.9$ & $ 0.041\pm 0.039$ & $-0.588\pm 0.000$ & $0.114\pm 0.004$ \\
     &      &  4 &\nodata& 2 & $ 5.6\pm  1.9$ & $ 0.006\pm 0.018$ & $-0.716\pm 0.289$ & $0.102\pm 0.035$ \\
\enddata
\tablecomments{(1) Redshift of input group (2) Early-type fraction of
  input group (3) Number of input galaxies brighter than $I_*+2$ (4)
  Input RS slope (5) Number of mock groups recovered (6) Mean number
  of recovered galaxies in RS (7) Mean recovered RS slope (8) Mean
  offset of RS effective color from input effective color (9) Mean
  scatter about fit RS}
\end{deluxetable*}

These tests indicate that we can detect rich groups via our detection
algorithm out to the highest lens redshifts in our sample, independent
of $f_{\rm e}$.  Groups with velocity dispersions of $\sim$300 km
s$^{-1}$ are detectable out to $z\approx 0.4$, although by $z=0.6$ the
detection rate is quite low ($\lesssim 50\%$).  Local Group analogs
are detectable at the $\sim$50\% level at low redshifts ($z=0.2$) if
$f_{\rm e}$ is close to one or to zero, although the former case is
unlikely to occur given that the early-type fractions of the poorest
groups are small \citep{Zabludoff1998}.  The detection rate of the
Local Group analogues at $z\geq 0.4$ is consistent with the
false-detection rate (see below).  We are therefore confident that we
can reliably detect the majority of galaxy groups with velocity
dispersions $\gtrsim 200$--300 km s$^{-1}$ out to $z=0.4$.  We can
detect poorer groups out to $z\approx 0.2$.

From the simulations we learn that, in the absence of morphological
information, the fit RS colors are systematically bluer than the input
value.  The reason is that some late-type galaxies (typically Sa's)
are included in the sample when the RS is fit.  For systems with
$f_{\rm e}$ near 1, the error is small.  In the groups with low
$f_{\rm e}$, we observe a broad peak in the effective color
distribution due to the late-type galaxies, and the estimated group
redshift is biased low (the magnitude of this bias is redshift
dependent).  We could lessen this systematic effect by including
morphological information in our galaxy selection; in future work we
will use galaxy profile fitting software \citep[e.g.,
\emph{GIM2D};][]{Simard2002} to reject disk-dominated systems from the
sample used to fit the RS.  Even without this morphological
information, we determine group redshifts with reasonable accuracy
(see \S\ref{spec.comp}).

We note that the individual fit RS slopes are very noisy and
imprecise, although the ensemble average is relatively accurate.  For
groups with $f_{\rm e}\approx 1$, the individual slope measurements are
much more accurate (Table~\ref{tab.artgrps}).  This suggests that the
accuracy of the slope determinations would be greatly improved by
using morphological data to exclude late-type galaxies from the RS
fit.

Finally, we note that the measured scatter about the RS is not a
reliable measure of the input scatter.  For groups with
$I_*+2.5\gtrsim 20$ ($z>0.25$), the photometric color errors dominate
the RS scatter.  By contrast, for groups with few RS members, the
paucity of red sequence galaxies results in an artificially low
scatter.  Only for richness 16 groups at low redshifts ($z\leq 0.25$)
is the recovered scatter comparable to the input scatter.

\emph{False positives} --- To determine the rate of false positives,
we run the group-finding algorithm on two lens fields, B1422 (for
$V-I$) and MG0751 (for $R-I$).  We assign each galaxy in the
photometric catalog a new, random position in the field, thereby
washing out any true groups and clusters.  We then search for peaks in
the effective color histogram, both within 0.05 mags of the lens
galaxy color (the typical scatter about the input color in the mock
RSs), simulating false positive detections of lens groups, and at
other colors, simulating false line-of-sight structures.  Fifty such
runs are performed in each field.

For the $V-I$ data, we record only two (i.e., 4\%) false positive
detections of lens galaxy groups and eight (16\%) false positive
line-of-sight groups.  For the $R-I$ data, 12 fields (24\%) have false
positive lens galaxy groups, and six (12\%) have false positive
line-of-sight groups.  The higher number of false positives in the
$R-I$ data arises mainly from the smaller color range spanned by the
catalog galaxies ($\sim$2 mags, compared to $\sim$4 mags in $V-I$).
Each color bin therefore contains more field galaxies, and the Poisson
fluctuations often exceed the group threshold.  The rate of false
positive lens groups in the MG0751 field is also enhanced by the fact
that the expected RS color at \zlens ($R-I=0.91$) lies near
the peak of the field galaxy color distribution, so shot noise from
the field population is even larger.  This latter problem vanishes for
$R$-band lens fields where $(R-I)_*\gtrsim 1$, or $\zlens\gtrsim
0.45$, as the field density of galaxies at these colors is much lower.

As a whole, false positives are not a serious concern for the lenses
in our sample with $V$ band imaging.  For lenses with $R$-band
imaging, we need to account for the false positive rate, especially
for lower-redshift lenses where the $R-I$ colors are similar to that
of the majority of background galaxies.  As will be shown in
\S\ref{sec.results.sample}, these false positives have a minimal
impact on our conclusions.

\subsection{Testing the Algorithm: Spectroscopic Confirmation\label{spec.comp}}

The ultimate test of the group finding algorithm is whether the
candidate groups can be confirmed as real, bound systems.  We have
obtained spectroscopic redshifts of $\sim 100$ galaxies in each of
eight lens fields (MG0751, BRI0952, PG1115, B1422, MG1654, PMN2004,
B2114, and HE2149); the identification of bound systems and the
analysis of their impact on lensing are discussed in
\citet{Momcheva2005}.  The group-finding algorithm detects eleven
groups in these eight fields, both at \zlens and elsewhere along the
line of sight.  Overall, the group finding algorithm works
exceptionally well, with ten of the eleven candidate groups confirmed
spectroscopically (Table \ref{tab.rsfits}).  The lone exception is the
RS at $z_{\rm RS}=0.20$ in the foreground of the PG1115 lens.  While
Figure 3 in \citet{Momcheva2005} shows a potential peak at this
redshift, there are too few velocities to determine if the peak is
real.  We also do not recover two spectroscopic line-of-sight groups
in our photometry.  The spectroscopic group in B1422 at $z=0.28$ is
blended with the lens group RS in color space, while the HE2149 group
at $z=0.27$ simply is not detected.

\begin{deluxetable*}{lcccccccc}
\tabletypesize{\footnotesize}
\tablewidth{0pt}
\tablecolumns{9}
\tablecaption{Parameters of significant candidate red sequences\label{tab.rsfits}}
\tablehead{\colhead{RS ID\tablenotemark{a}} & \colhead{$z_{\rm RS}$} & \colhead{$z_{\rm spec}$} & \colhead{$I_*$} &
  \colhead{$(V-I)_*$} & \colhead{$(R-I)_*$} & \colhead{RS Slope} &  
  \colhead{$N_{\rm RS}$} & \colhead{RS Scatter}}
\startdata
B0712 : GROUP 1 & 0.63 &\nodata & 19.903 & \nodata & $1.361\pm 0.029$ & $0.115\pm 0.029$ &  9.2 & 0.088 \\
MG0751: GROUP 1 & 0.34 & 0.3502 & 18.229 & \nodata & $0.900\pm 0.051$ & $0.093\pm 0.032$ &  5.8 & 0.044 \\
MG0751: GROUP 2 & 0.48 & 0.5605 & 19.149 & \nodata & $1.087\pm 0.037$ & $0.121\pm 0.023$ & 13.1 & 0.047 \\
FBQS0951: GROUP 1&0.16 & \nodata& 16.461 & $1.434\pm 0.060$ & \nodata & $0.052\pm 0.042$ &  5.6 & 0.038 \\
FBQS0951: GROUP 2&0.27 & 0.24\tablenotemark{b} & 17.675 & $1.770\pm 0.041$ & \nodata & $0.062\pm 0.032$ &  5.8 & 0.047 \\
FBQS0951: GROUP 3&0.43 &\nodata & 18.844 & $2.349\pm 0.037$ & \nodata & $0.034\pm 0.024$ &  6.4 & 0.051 \\
BRI0952: GROUP 1& 0.37 & 0.4220 & 18.441 & \nodata & $0.938\pm 0.057$ & $0.059\pm 0.048$ &  8.4 & 0.020 \\
PG1115: GROUP 3 & 0.26 &\nodata & 17.587 & $1.732\pm 0.162$ & \nodata & $0.063\pm 0.081$ &  4.7 & 0.038 \\
PG1115: GROUP 1 & 0.30 & 0.3101 & 17.924 & $1.895\pm 0.045$ & \nodata & $0.078\pm 0.017$ &  3.4 & 0.031 \\
PG1115: GROUP 2 & 0.41 & 0.4859 & 18.713 & $2.297\pm 0.077$ & \nodata & $0.116\pm 0.042$ &  3.4 & 0.080 \\
RXJ1131: GROUP 1& 0.19 &$\approx 0.104$\tablenotemark{c} & 16.864 & $1.508\pm 0.024$ & \nodata & $0.059\pm 0.020$ & 25.6 & 0.045 \\
RXJ1131: GROUP 2& 0.29 & 0.295\tablenotemark{b} & 17.844 & $1.852\pm 0.077$ & \nodata & $0.071\pm 0.046$ &  6.2 & 0.056 \\
B1422: GROUP 1  & 0.30 & 0.3387 & 17.924 & $1.895\pm 0.049$ & \nodata & $0.104\pm 0.031$ &  6.0 & 0.054 \\
B1600: GROUP 1  & 0.50 &\nodata & 19.262 & \nodata & $1.123\pm 0.064$ & $0.086\pm 0.041$ & 12.1 & 0.035 \\
MG1654: GROUP 1 & 0.22 & 0.2527 & 17.204 & $1.596\pm 0.046$ & \nodata & $0.079\pm 0.033$ &  4.8 & 0.021 \\
B2114: GROUP 1  & 0.26 & 0.3144 & 17.587 & \nodata & $0.792\pm 0.058$ & $0.055\pm 0.030$ &  7.0 & 0.062 \\
HE2149: GROUP 3 & 0.40 & 0.4465 & 18.647 & \nodata & $0.968\pm 0.060$ & $0.047\pm 0.036$ &  5.6 & 0.052 \\
HE2149: GROUP 4 & 0.59 & 0.603  & 19.720 & \nodata & $1.288\pm 0.068$ & $0.086\pm 0.042$ &  5.3 & 0.041 \\
\enddata
\tablenotetext{a}{Matched to \citet{Momcheva2005} where applicable}
\tablenotetext{b}{No group data in \citet{Momcheva2005}; lens galaxy redshift from 
Table~\ref{tab.sample}}
\tablenotetext{c}{Redshift of BRSG from Las Campanas Redshift Survey \citep{Shectman1996}}
\tablecomments{The IAU-approved naming convention for these groups is
  Lens Name: MWKZ GROUP N, where N is the group ID; e.g., CLASS
  B0712+472: MWKZ GROUP 1}
\end{deluxetable*}

The photometric redshifts tend to be slightly lower than the
spectroscopic redshifts ($\overline{\Delta z} = -0.02\pm0.04$),
because the group finding algorithm tends to return colors that are
slightly bluer than the true RS, as discussed in
\S\ref{sec.algorithm.tests}.  This offset is small, which shows that
our photometric group finding algorithm obtains reliable redshifts for
detected groups of galaxies.

\begin{figure}
\plotone{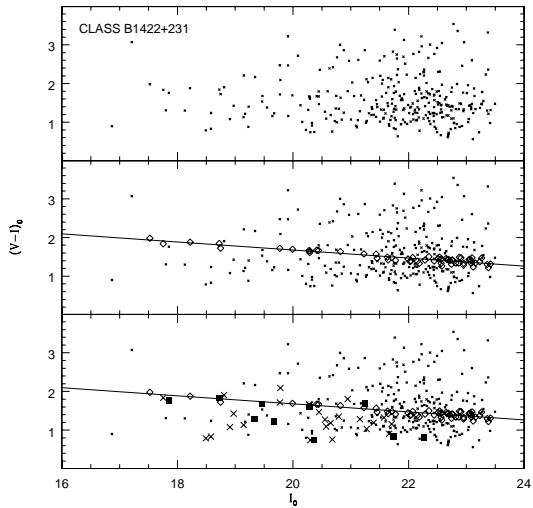}
\caption{Comparison of RS-finding algorithm with spectroscopic data
  for B1422.  \emph{Top panel}: Extinction-corrected color-magnitude
  diagram for all galaxies (small crosses) within a projected \rvir
  radius surrounding the centroid of the B1422 group.  \emph{Middle
  panel}: Same as top panel, but with best-fitting RS (solid line) and
  selected candidate RS-member galaxies (open diamonds) shown.
  \emph{Bottom panel}: Same as middle panel, but with
  spectroscopically confirmed group members from \citet{Momcheva2005}
  (filled squares, including galaxies from beyond \rvir) and confirmed
  non-members (large crosses) shown.  This figure shows that the fit
  RS agrees with the red edge of the envelope defined by the
  spectroscopically-confirmed group members.
  \label{fig.b1422_speccomp}}
\end{figure}

\begin{figure}
\plotone{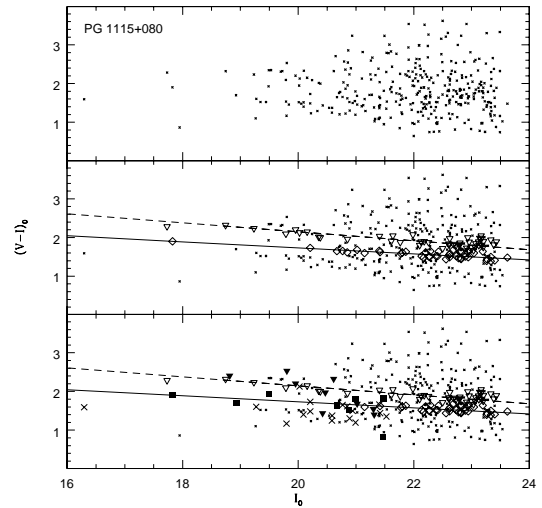}
\caption{Same as Figure \ref{fig.b1422_speccomp}, but for PG1115. Open
  triangles (center and bottom panels) indicate the candidate RS
  member galaxies for the background group; filled triangles (bottom
  panel) indicate spectroscopic members of the background group. The
  dashed line indicates the fit RS for the background group.  This
  figure illustrates the ability of the red-sequence finding algorithm
  to detect multiple groups in a given field.
\label{fig.pg1115_speccomp}} 
\end{figure}

 Figure \ref{fig.b1422_speccomp} shows the spectroscopic group members
and non-members in comparison with the fit RS for B1422.  The fit RS
agrees with the red edge of the envelope defined by the group members.
Figure \ref{fig.pg1115_speccomp} shows the spectroscopic group members
and fit RSs for both the lens group and a background group in PG1115.
The background group in PG1115 has a significantly higher
spectroscopic redshift ($z_{\rm sp}=0.486$) than the photometric RS
redshift ($z_{\rm RS}=0.41$), suggesting that this group may have a
low $f_{\rm e}$; indeed, \citet{Momcheva2005} find that six of the ten
confirmed group members have emission lines.  Both of these figures
illustrate the success of our RS finding technique.

We define a red-sequence richness parameter $N_{\rm RS}$ to be the
number of selected RS galaxies brighter than $I_*+2.5$ over and above
the normalized background. With two-band photometry alone, this is the
best richness measure we can make. We compare $N_{\rm RS}$ with the
measured velocity dispersions from \citet{Momcheva2005}.  Contrary to
expectations, we find no significant correlation. The explanation may
be that variable early-type fractions and statistical background noise
add significant scatter \citep[see also the large scatter in
$N_{*666}$ versus $\sigma$ for 2MASS galaxy clusters with $\sigma
\lesssim 500$ km s$^{-1}$; ][]{Kochanek2003}.  The full
sample of groups from our complete survey will overcome this issue by
spanning a much larger dynamic range, making the comparison between
$N_{\rm RS}$ and $\sigma$ more meaningful.

\section{Results and Discussion\label{sec.results}}

\subsection{The Detected Red Sequence Sample}
In total, eighteen significant candidate RSs are detected in twelve
lens fields in the present sample.  The fit RS parameters are
presented in Table~\ref{tab.rsfits}, and the photometry of the
selected galaxies is presented in Table \ref{tab.b0712.1} (available
electronically).

\begin{deluxetable*}{lcrcccccc}
\tablewidth{0pt}
\tablecolumns{9}
\tablecaption{Photometry of candidate red sequence galaxies \label{tab.b0712.1}}
\tablehead{\colhead{Lens} & \colhead{Group} & \colhead{Galaxy} & \colhead{RA} & 
\colhead{Dec} & \colhead {$I$} & \colhead {$\delta I$} & \colhead{$R-I$} & 
\colhead{$\delta(R-I)$}\\
& \colhead{ID} & \colhead{ID} & & & & & &}
\startdata
B0712  & 1 &  8471 & 07 16 04.35 &  47 08 59.2 & 18.61 & 0.03 &  1.60 & 0.03\\
B0712  & 1 &  9253 & 07 15 56.86 &  47 09 23.7 & 20.80 & 0.03 &  1.45 & 0.07\\
B0712  & 1 &  9109 & 07 15 58.92 &  47 08 52.3 & 20.82 & 0.03 &  1.45 & 0.01\\
B0712  & 1 &  9412 & 07 15 55.16 &  47 09 53.5 & 21.21 & 0.05 &  1.27 & 0.11\\
B0712  & 1 &  8204 & 07 16 09.09 &  47 10 04.5 & 21.28 & 0.05 &  1.10 & 0.20\\
\enddata
\tablecomments{(1) Full table is available electronically.  (2) The 
IAU-approved naming convention for individual galaxies
  is: Lens Name: MWKZ GAL NNNN, where NNNN is the galaxy ID above;
  e.g., B0712+472: MWKZ GAL 8471. (3) Units of right ascension are
  hours, minutes, and seconds, and units of declination are degrees,
  arcminutes, and arcseconds (J2000.0).}  
  
\end{deluxetable*}

Although some final centroids lie outside \rvir of the lens, we initially
detected each RS peak within \rvir of the lens
galaxy.  It is certain that the wide-field images contain additional
structures outside of this initial search radius; no attempt has been
made so far to locate such structures.

The average RS slope, $\Delta (V - I) / I$ or $\Delta (R - I) / I$, for
all 18 candidate groups is $0.076\pm 0.025$.  This scatter is
comparable to that measured in the mock groups
(Table~\ref{tab.artgrps}), so there is no evidence that the RS slope
is anything but universal at these moderate redshifts.  We note that
these slopes are not corrected for redshift/evolutionary effects.
Such a correction would require a detailed understanding of the origin
of the RS and its redshift evolution.

\subsubsection{The Fraction of Lenses in Complex Environments
  \label{sec.results.sample}} 

Of the twelve lens systems in this paper, eight have significant red
sequences within $\Delta z=0.06$ of the 
lens redshift. Of these systems, six were known or
suspected previously to have a group at the lens redshift due to the
need for shear in the lens models or to an overdensity of galaxies on the
sky surrounding the lens. The two new systems are FBQS0951 ($z_{\rm
RS} = 0.27$) and B2114 ($z_{\rm RS} = 0.26$); the B2114 group has been
confirmed spectroscopically \citep{Momcheva2005}.

\begin{figure*}
\includegraphics[width=5in]{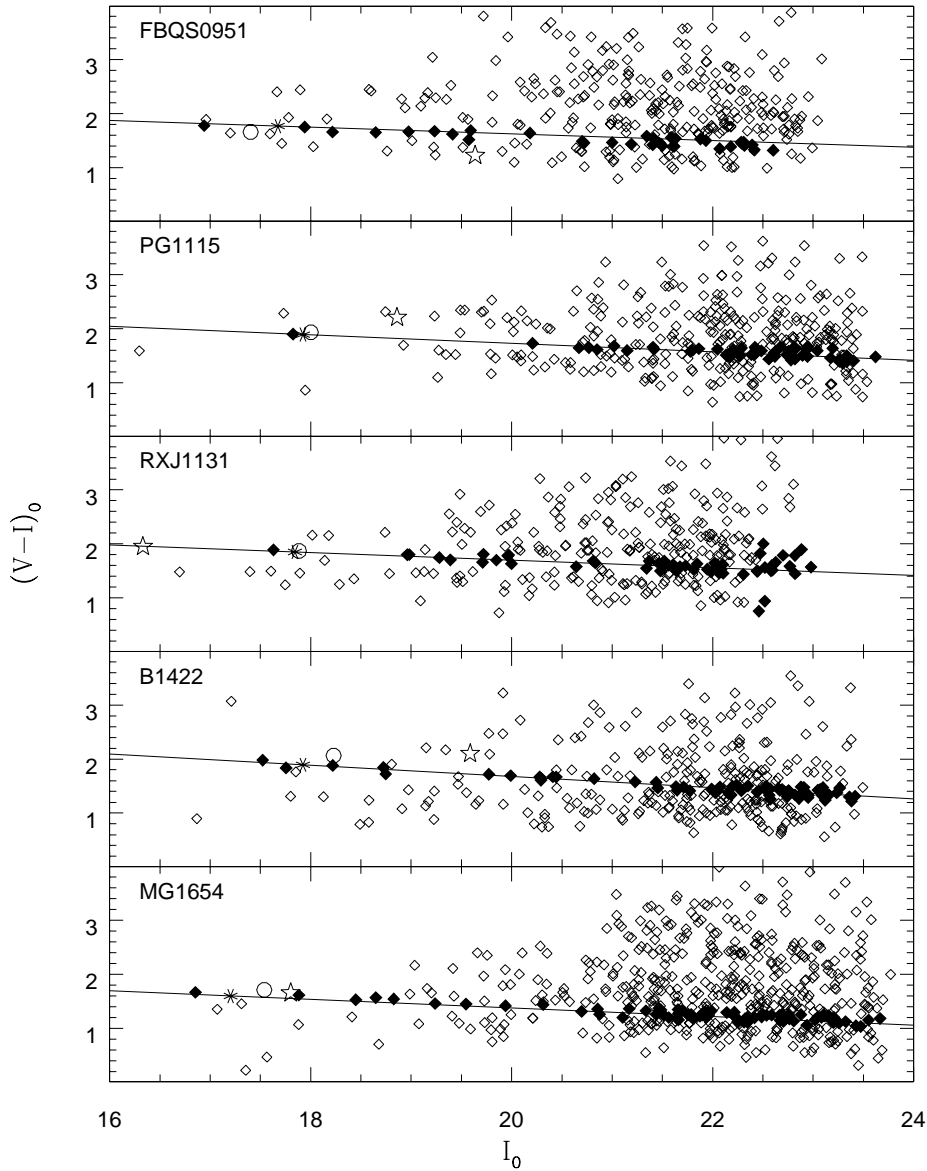}
\caption{Color-magnitude diagrams and red sequences at the lens
  redshift. Each panel presents an individual lens. Open diamonds show all
  galaxies within \rvir ($\equiv 500h^{-1}$ kpc) of
  the group centroid (or of the lens galaxy for those lenses with no
  detected red sequence).  Selected candidate red
  sequence galaxies are shown as filled diamonds, along with the
  best-fitting red sequence for galaxies with $I\leq I_*+2.5$ (solid
  line).  The lens galaxies are denoted by large, open stars.
  Asterisks indicate the location of an $L_*$ galaxy along the fit red
  sequence; open circles show the location of an $L_*$ galaxy at
  \zlens.  For the lenses with no detected red sequence (B0712, B1600,
  HE2149, and PMN2004), a dotted line indicates a nominal red sequence
  with a slope of 0.09 at \zlens.  Eight of the twelve lenses have a
  red sequence at the lens redshift. \label{fig.rs_at_lensz}}
\end{figure*}

Figure \ref{fig.rs_at_lensz} shows the CMDs for each of the twelve
lens fields.  In the cases where an RS is detected at the lens
redshift, the fit RS and the selected RS galaxies are shown.  Where no
RS is detected at the lens redshift, a nominal RS for the lens
redshift is shown with a slope of $\beta_V=0.09$.  In many cases, the
detected RS is not apparent to the unaided eye, but our spectroscopy
has established that these are physical systems, showing the power of
our algorithm.

\begin{figure*}
\includegraphics[width=6in]{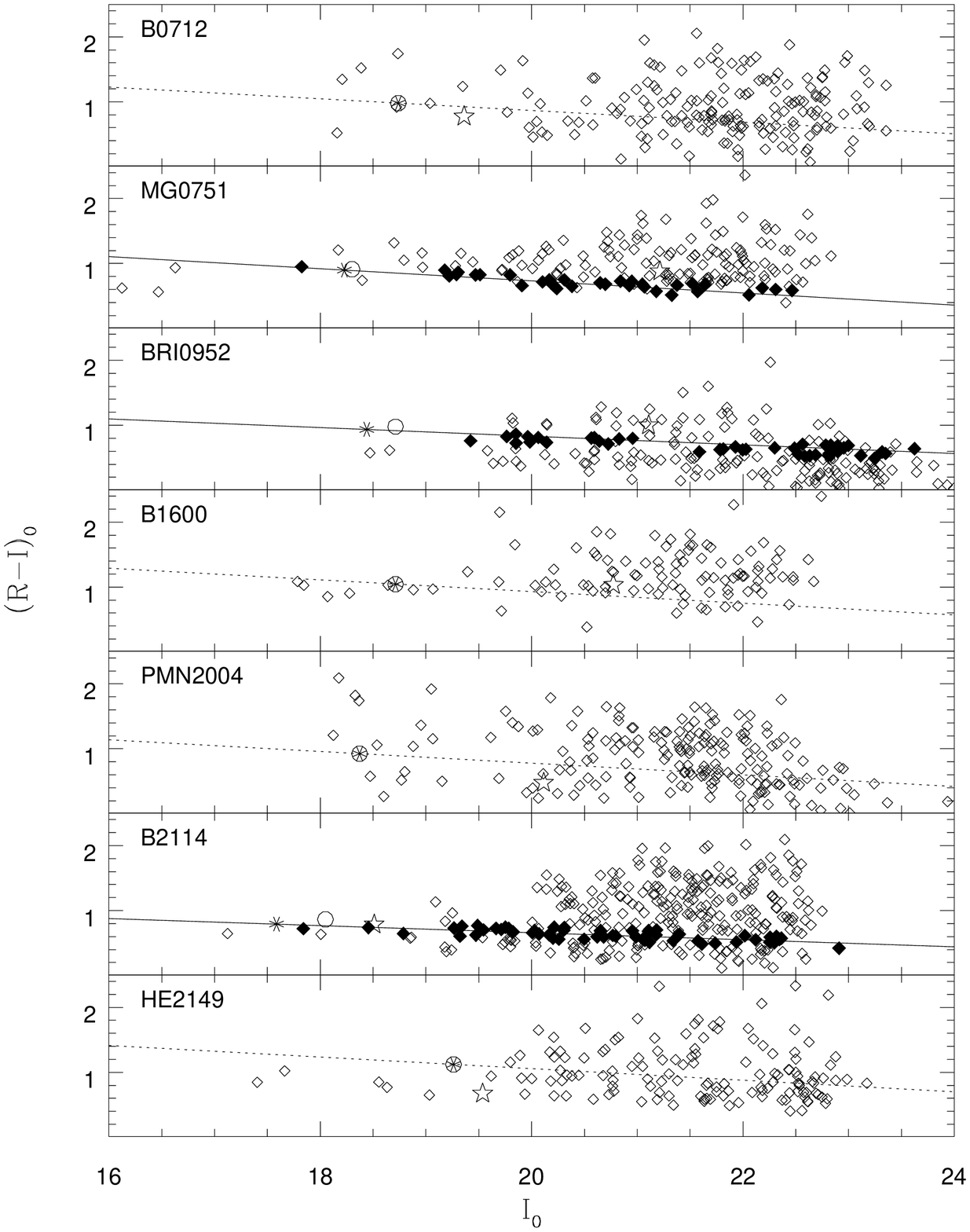}\\
\centerline{Fig.~\ref{fig.rs_at_lensz} (cont.)}
\end{figure*}

\begin{figure*}
\plotone{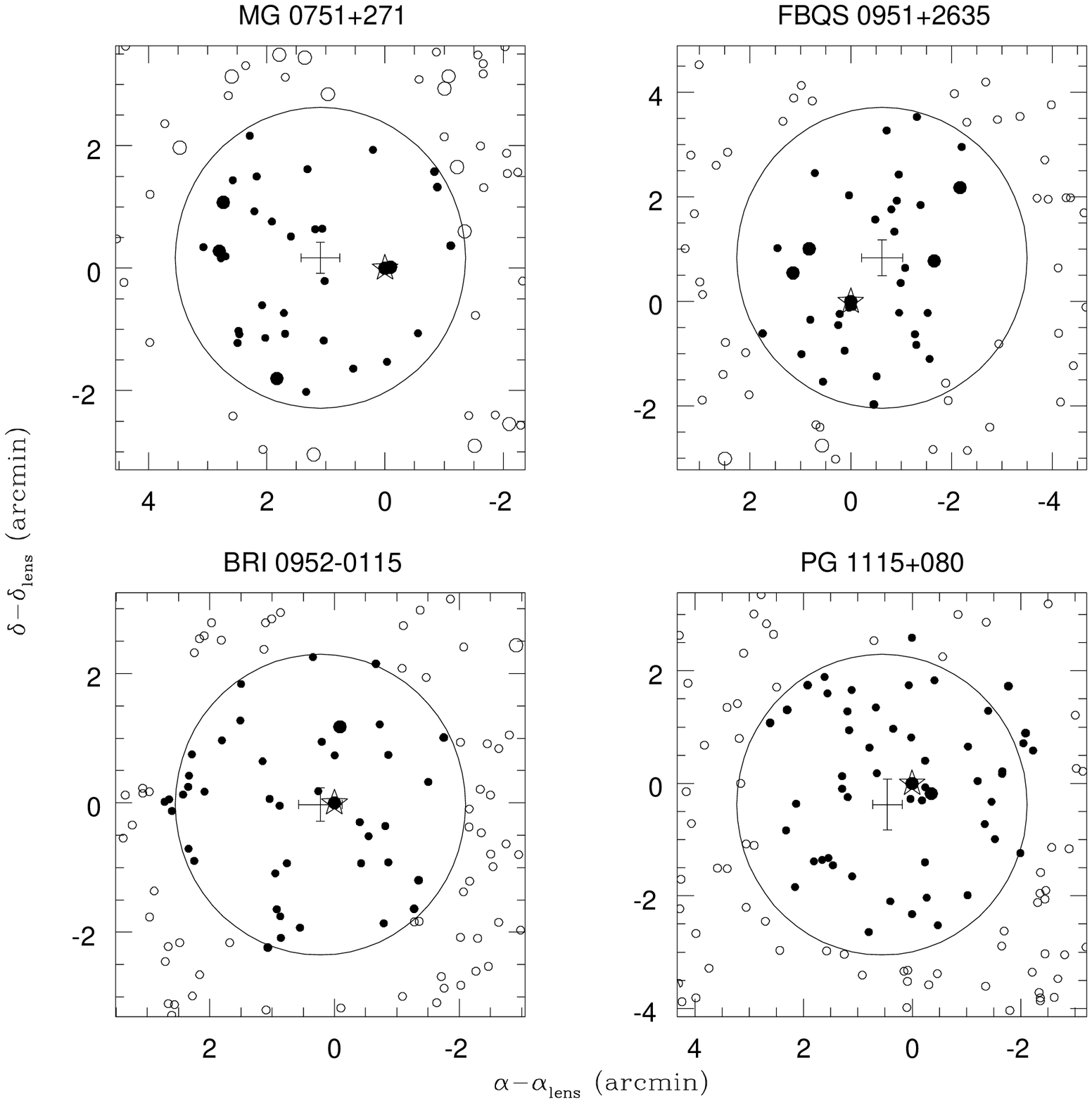}
\caption{Sky maps of red sequence members of candidate lens groups.
  Point sizes are scaled by galaxy magnitude relative
  to $I_*$.  The open star at each map's origin marks the lens
  position, and the error bars denote the positions of the unweighted
  red sequence centroids and their 1$\sigma$ errors.  The circle shows one 
  projected group virial radius ($\equiv 500 h^{-1}\,{\rm kpc}$)
  at the red sequence redshift centered on the
  group centroid. These maps show that, in many cases, the lens galaxy is
  not consistent with the group centroid and is not the brightest
  galaxy in the red sequence. \label{fig.maps}}
\end{figure*}

Figure \ref{fig.maps} presents sky maps of galaxies in each of the
eight detected RSs near \zlens.  In many cases, the galaxies appear to
be clustered, as would be expected in a physical group.  In a few
cases, such as BRI0952, there appears to be little, if any clustering
of galaxies, yet \citet{Momcheva2005} do find a physical group around
BRI0952 at the lens redshift.

\begin{figure*}
\plotone{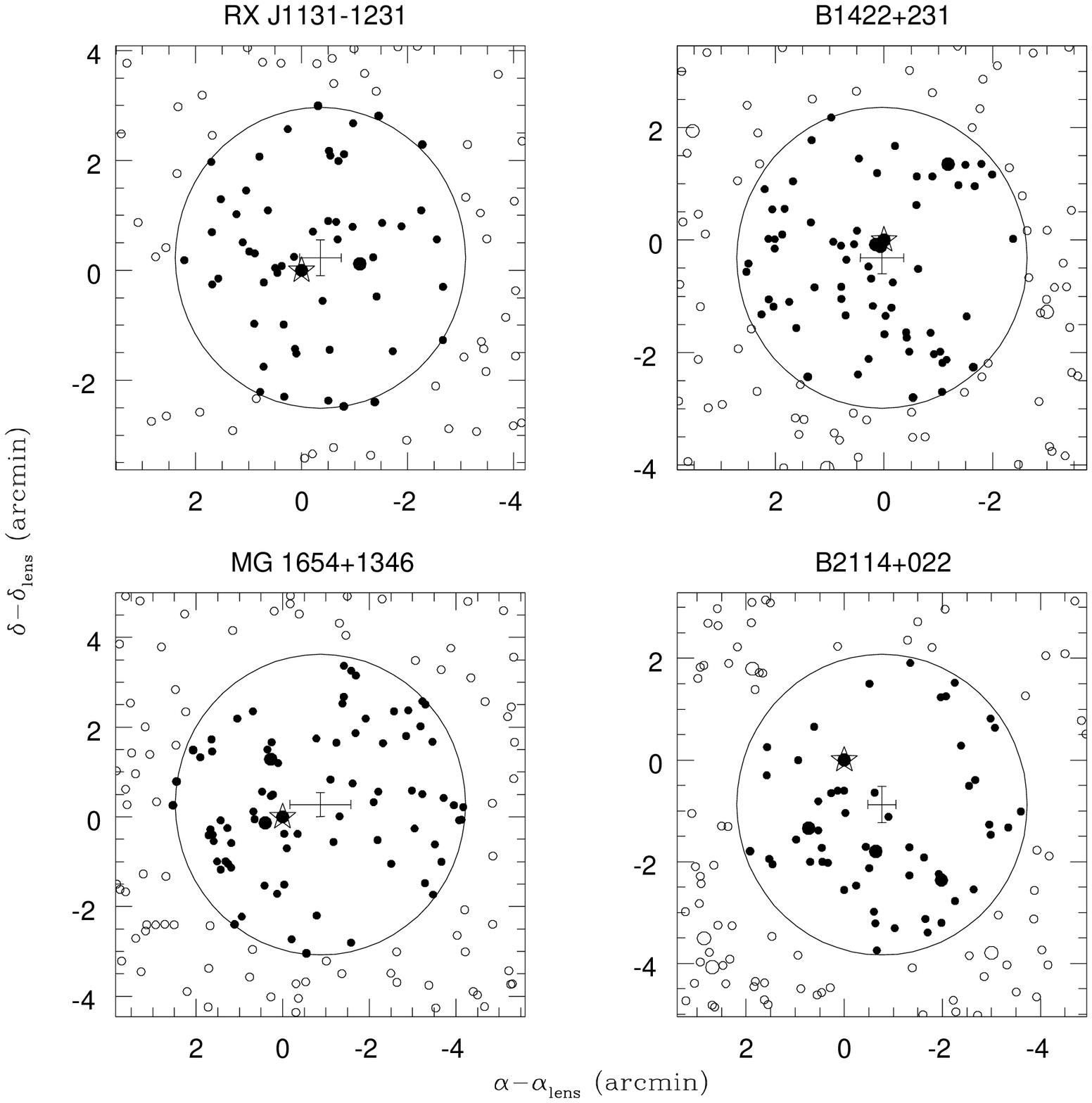}
\centerline{Fig.~\ref{fig.maps} (cont.)}
\end{figure*}

The sky maps also show that the lens galaxy (indicated by a star at
the origin in each plot) is not necessarily at the center of the
group.  Table~\ref{tab.rscent} presents the raw and
luminosity-weighted projected centroids for each candidate RS, plus
(where available) the offset between the raw RS centroid and the
centroid of spectroscopically confirmed group members from
\citet{Momcheva2005}.  For four of the six groups, the centroids agree
to within $2\sigma$, while for the other two (MG0751: MWKZ Group 1 and
BRI0952: MWKZ Group 1), the photometric and spectroscopic centroids
disagree.  In these cases, the disagreement could be due to
contamination of the photometric sample by non-member galaxies, to
incompleteness in the spectroscopic sample, or to some combination of
the two.  We are analyzing additional spectroscopic data for these
groups, which should improve the spectroscopic completeness and hence
the centroid accuracy.

\begin{deluxetable*}{lccccccc}
\tabletypesize{\footnotesize}
\tablewidth{0pt}
\tablecolumns{6}
\tablecaption{Centroids of candidate red sequences\label{tab.rscent}}
\tablehead{\colhead{RS ID} & 
  \multicolumn{2}{c}{Raw Centroid} & \multicolumn{2}{c}{$L$-weighted Centroid} & 
  \colhead{Offset from \citet{Momcheva2005}} \\ \cline{2-3} \cline{4-5}
& \colhead{$\alpha$} & \colhead{$\delta$} &
  \colhead{$\alpha$} & \colhead{$\delta$} & \colhead{(\arcsec)}}
\startdata
B0712: GROUP 1  & 07 16 03 &  $\phantom{-}$47 09 25 & 07 16 03 &  $\phantom{-}$47 09 14 &  \nodata \\
MG0751: GROUP 1 & 07 51 48 &  $\phantom{-}$27 16 42 & 07 51 47 &  $\phantom{-}$27 16 45 & $ 101\pm 20$ \\
MG0751: GROUP 2 & 07 51 44 &  $\phantom{-}$27 17 10 & 07 51 44 &  $\phantom{-}$27 17 09 & \nodata \\
FBQS0951: GROUP 1& 09 51 22 &  $\phantom{-}$26 35 01 & 09 51 23 &  $\phantom{-}$26 35 10 & \nodata \\
FBQS0951: GROUP 2& 09 51 20 & $\phantom{-}$26 36 19 & 09 51 19 &  $\phantom{-}$26 36 19 & \nodata \\
FBQS0951: GROUP 3& 09 51 22 &  $\phantom{-}$26 36 01 & 09 51 23 &  $\phantom{-}$26 35 51 & \nodata \\
BRI0952: GROUP 1& 09 55 03 &  $-$01 30 02 & 09 55 03 & $-$01 30 01 &  $ 107\pm 41$ \\
PG1115: GROUP 3 & 11 18 16 &  $\phantom{-}$07 43 27 & 11 18 16 &  $\phantom{-}$07 43 12 & \nodata \\
PG1115: GROUP 1 & 11 18 18 &  $\phantom{-}$07 45 27 & 11 18 17 &  $\phantom{-}$07 45 47 &  $  20\pm 22$ \\
PG1115: GROUP 2 & 11 18 18 &  $\phantom{-}$07 46 03 & 11 18 19 &  $\phantom{-}$07 46 24 & \nodata \\
RXJ1131: GROUP 1& 11 32 00 & $-$12 31 48 & 11 32 03 & $-$12 32 03 & \nodata \\
RXJ1131: GROUP 2& 11 31 50 & $-$12 31 45 & 11 31 49 & $-$12 31 45 & \nodata \\
B1422:  GROUP 1 & 14 24 38 &  $\phantom{-}$22 55 54 & 14 24 37 &  $\phantom{-}$22 56 15 &  $  40\pm 24$ \\
B1600:  GROUP 1 & 16 01 45 &  $\phantom{-}$43 16 48 & 16 01 45 &  $\phantom{-}$43 16 44 & \nodata \\
MG1654: GROUP 1 & 16 54 37 &  $\phantom{-}$13 46 38 & 16 54 40 &  $\phantom{-}$13 46 39 &  $  54\pm 39$ \\
B2114:  GROUP 1 & 21 16 47 &  $\phantom{-}$02 24 54 & 21 16 47 &  $\phantom{-}$02 24 41 &  $  68\pm 40$ \\
HE2149: GROUP 3 & 21 52 08 & $-$27 32 17 & 21 52 06 & $-$27 32 09 & \nodata \\
HE2149: GROUP 4 & 21 52 06 & $-$27 31 31 & 21 52 05 & $-$27 31 36 & \nodata \\
\enddata
\tablecomments{Units of right ascension are hours, minutes, and
seconds, and units of declination are degrees, arcminutes, and
arcseconds (J2000.0).}
\end{deluxetable*}

Among the eight lenses with an RS at the lens redshift, the lens galaxy
position is within 1$\sigma$ of the projected RS centroid for two
systems, and within 3$\sigma$ for two more.  Stated another way, in
four cases the lens galaxy is \emph{not} consistent with lying at the
centroid (even accounting for the centroid uncertainties).  This
strongly suggests that lenses do not necessarily occupy the center of
the local mass distribution, which is an important realization for
lens modeling \citep[see \S\ref{sec.brsg} and][]{Momcheva2005}.

\subsubsection{The Fraction of Lenses with Line-of-Sight Structures}
Seven of the twelve lens systems have RSs that are fore or aft of the
lens along the line of sight; prior to this study, only one of these
groups (B0712) was known to have a line-of-sight group, at $z\approx
0.29$ \citep{Fassnacht2002}.  This group is not detected here, although
we do detect a background group at $z\approx 0.67$.  Two of the
line-of-sight systems (MG0751: MWKZ Group 2 and RXJ1131: MWKZ Group 1)
have very rich RSs indicative of a rich group or even a cluster of
galaxies, while the remaining six are more suggestive of poor groups.

\begin{figure*}
\includegraphics[scale=0.67]{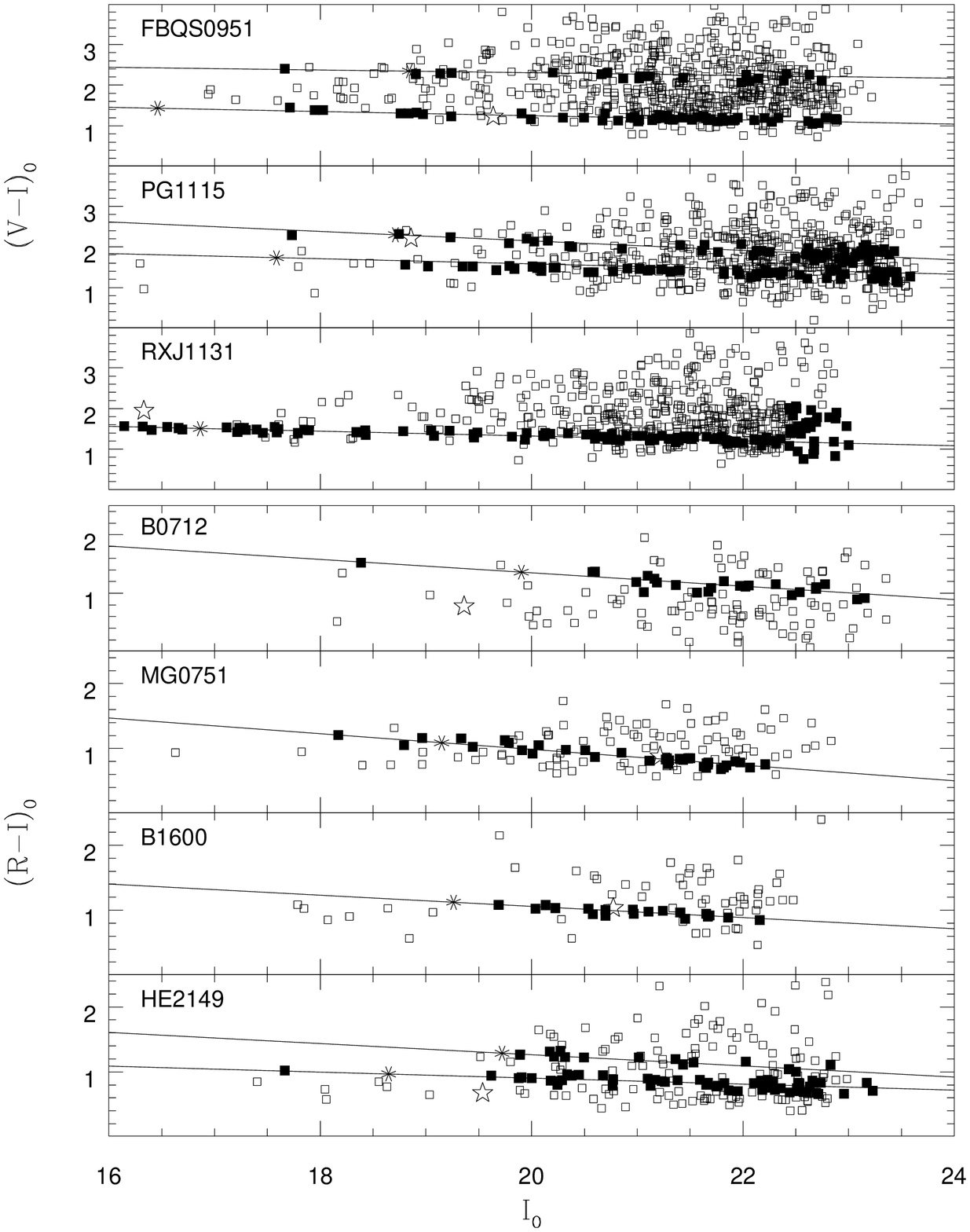}
\caption{Color-magnitude diagrams for candidate line-of-sight red
  sequences.  Symbols are as in Fig.~\ref{fig.rs_at_lensz}. Ten
  line-of-sight lenses are found in seven (of twelve total) lens
  fields.  Although to the naked eye many of these red sequences are
  not obvious, our spectroscopic observations of
  HE2149, MG0751, and PG1115 confirm the
  line-of-sight groups in HE2149 and MG0751 and the background group
  in PG1115 (Momcheva et al.~2006).  The low-$z$ line-of-sight
  candidate in PG1115 was not detected spectroscopically. The
  apparently steeper red-sequence slopes for the lower four panels are
  due to the smaller vertical color range in $R-I$.\label{fig.rs_los}}
\end{figure*}

Figure \ref{fig.rs_los} shows the CMDs for all candidate line-of-sight
groups.  We observe a rich, tight RS in RXJ1131, indicative of a
foreground cluster of galaxies; indeed, two of these galaxies have
redshift measurements of $z\approx 0.10$ from the Las Campanas
Redshift Survey \citep{Shectman1996}, and there is extended X-ray
emission consistent with their position (C.~Kochanek, private
communication).  Some of the candidate RSs appear dubious, such as the
background groups in MG0751 and HE2149. Once again, however, our
spectroscopic follow-up confirms the existence of these groups.

Based on the simulations presented in \S\ref{sec.algorithm.tests}, we
expect about one false positive (either line-of-sight or at \zlens)
among the five lens fields observed in the $V$-band and two to three
false positives from the seven lens fields with $R$-band data.

\subsection{Lens Galaxy Environments and Shear}

It is interesting to consider whether observed lens galaxy
environments can explain the external tidal shears required in lens
models.  Shear is created by an asymmetric distribution of mass around
the lens galaxy; a simple way to get that is to have the lens galaxy
offset from the centroid of the group.  Historically, several attempts
to compare the shear required by lens models with the distribution of
galaxies within $\sim 140$ kpc of the lens have yielded mixed results.
In two quadruple-image lenses (PG1115 and B1422), the model shear
appears to be consistent with the distribution of galaxies near the
lens \citep{Hogg1994,Schechter1997,Keeton1997,Kundic1997,Kundic1997b}.
In contrast, in the Einstein ring MG0751, the shear angle is {\em not}
consistent with the immediate lens environment \citep{Lehar1997} Also,
in a sample of 10 double-image lenses, \citet{Lehar2000} do not find
any strong correlation between model shears and close environments.

We can improve upon the previous studies in three ways.  First, we
have obtained a larger sample of lens environments.  Second, we have
measured lens environments over a larger area on the sky.  Third, we
have identified enough group member galaxies to make a reliable
measurement of the group centroid position.  One limitation of our
present analysis is that, without measured velocity dispersions, we
cannot estimate the amplitude of the shear contributed by each lens
environment.  \citep[See][ for an analysis of shear amplitudes for a
smaller sample of lenses.]{Momcheva2005} Nevertheless, we can still
compare the shear position angle with the position angle of the group
centroid (relative to the lens galaxy) and determine whether the
observed environments are at least consistent with the model shears.

Table \ref{tab.lensprops} gives the position angles for the
observed group centroid and the lens model shear, where available.  In
all six cases where both are available, the centroid and shear
position angles are remarkably consistent (albeit with relatively
large centroid errorbars in a few cases).  Perhaps the most
interesting system is MG0751: while the immediate lens environment cannot
explain the model shear \citep{Lehar1997}, we now see that the larger
environment can.  This example illustrates why it is important to
observe lens environments over an area large enough to include the
full virial extent of a group at the lens redshift.  It will be
interesting to reconsider this question in more detail, after
measuring velocity dispersions so that we may estimate shear
amplitudes as well as position angles, and after making new lens
models of all the systems.  Still, it is very encouraging to see new
evidence that the observed lens environments can explain the shears
required by lens models.

\begin{deluxetable*}{lrrrrccccc}
\tabletypesize{\footnotesize}
\tablewidth{0pt}
\tablecolumns{10}
\tablecaption{Comparison of group centroids to lens properties\label{tab.lensprops}}
\tablehead{\colhead{Lens} & \multicolumn{2}{c}{Raw Centroid} & 
\multicolumn{2}{c}{$L$-weighted Centroid} & \colhead{Offset} & \multicolumn{2}{c}{Position Angle} & Shear & Reference \\ 
\cline{2-3} \cline{4-5} \cline{7-8} & \colhead{$\Delta\alpha$} & \colhead{$\Delta\delta$} & \colhead{$\Delta\alpha$} & 
\colhead{$\Delta\delta$} & \colhead{(kpc)} & \colhead{Raw Centroid} & \colhead{$L$-weighted} & PA & }
\startdata
MG0751 & $ 90\pm 18$ & $  10\pm 15$ & $ 76\pm 26$ & $  13\pm 14$ & $440\pm  90$ & $  84\pm 10$ & $  80\pm 14$ & $  64$ & (1) \\
FBQS0951&$-37\pm 22$ & $  65\pm 20$ & $-53\pm 32$ & $  65\pm 16$ & $310\pm  90$ & $ -30\pm 17$ & $ -39\pm 19$ & \nodata & \nodata \\
BRI0952& $ 47\pm 21$ & $   4\pm 16$ & $ 48\pm 18$ & $   6\pm 18$ & $240\pm 110$ & $  85\pm 27$ & $  83\pm 24$ & $  66$ & (2) \\
PG1115 & $ 15\pm 16$ & $ -32\pm 27$ & $ -4\pm 22$ & $ -12\pm 20$ & $160\pm 110$ & $ 155\pm 47$ & $-162\pm 78$ & $-113$ & (3) \\
RXJ1131& $-21\pm 23$ & $  14\pm 20$ & $-35\pm 21$ & $  13\pm 15$ & $110\pm 100$ & $ -56\pm 58$ & $ -70\pm 35$ & $ -80$ & (4) \\
B1422  & $  2\pm 22$ & $  -7\pm 17$ & $-17\pm 18$ & $  15\pm 17$ & $ 30\pm  80$ & $ 164\pm 86$ & $ -49\pm 54$ & $ -53$ & (4) \\
MG1654 & $-76\pm 41$ & $  16\pm 16$ & $-23\pm 37$ & $  17\pm 18$ & $280\pm 140$ & $ -78\pm 28$ & $ -53\pm 63$ & $ -80$ & (5) \\
B2114  & $-58\pm 17$ & $ -53\pm 21$ & $-52\pm 18$ & $ -66\pm 22$ & $320\pm  80$ & $-132\pm 15$ & $-142\pm 14$ & \nodata & \nodata \\
\enddata
\tablecomments{Units of $\Delta\alpha$ and $\Delta\delta$ are arcseconds (J2000.0). PA measured
East of North.  Except for PG1115, shear PAs are defined modulo 180 degrees.}
\tablerefs{(1) \citealt{Lehar1997}, (2) \citealt{Lehar2000}, (3) \citealt{Impey1998}, 
(4) \citealt{Keeton2003}, (5) \citealt{Kochanek1995}}
\end{deluxetable*}

\subsection{Lens Galaxy Magnitude Distribution\label{sec.lensmagdist}}

The sample of lenses allows us to examine the
distribution of lens galaxy magnitudes.  While bright (i.e., massive)
galaxies have large lensing cross-sections, there are many more
faint galaxies in groups \citep[e.g.,][]{Keeton2000}. Taken together, these effects lead one to expect the typical lens to be $\sim L_*$
\citep{Kochanek2000,Kochanek2001}, though it is still unclear how faint (and
dwarf-like) lenses can be. In our sample, \emph{ten} of the eleven
lens galaxies with redshift measurements are fainter than $I_*$ (see
Table \ref{tab.lensmags}). In order to determine whether our sample
includes some unknown bias toward faint lenses (although one would
expect that any bias would be toward bright, massive galaxies), we
calculate the expected distribution of lens galaxy magnitudes based on
theoretical expectations and a minimal number of reasonable
assumptions.  We then compare this theoretical distribution with the
observed distribution in our sample and a larger sample of
gravitational lens galaxies from \citet{Rusin2003}.  

\begin{deluxetable}{lcc}
\tablewidth{0pt}
\tablecolumns{8}
\tablecaption{Lens Galaxy Magnitudes\label{tab.lensmags}}
\tablehead{\colhead{Lens} & \colhead{$I_{\rm lens}$} & \colhead{$I_{0,\rm lens} - I_*$} }
\startdata
B0712    & $19.58\pm 0.07$ & $\phantom{-}0.64$ \\
MG0751   & $21.28\pm 0.64$ & $\phantom{-}3.05$ \\
FBQS0951 & $19.68\pm 0.23$ & $\phantom{-}2.01$ \\
BRI0952  & $21.23\pm 0.04$ & $\phantom{-}2.79$ \\
PG1115   & $18.94\pm 0.02$ & $\phantom{-}1.02$ \\
RXJ1131  & $16.6\phantom{0}\pm 0.3\phantom{0}$ & $-1.3\phantom{0}$ \\
B1422    & $19.68\pm 0.25$ & $\phantom{-}1.76$ \\
B1600    & $20.80\pm 0.41$ & $\phantom{-}1.54$ \\
MG1654   & $17.92\pm 0.02$ & $\phantom{-}0.72$ \\
PMN2004  & $20.51\pm 0.02$ & \nodata \\
B2114    & $18.65\pm 0.23$ & $\phantom{-}1.06$ \\
HE2149   & $19.60\pm 0.03$ & $\phantom{-}0.87$ \\
\enddata
\tablecomments{Except for PMN2004, all lens galaxy magnitudes are from the literature (see text).}
\end{deluxetable}

Lensing's sensitivity to mass means that the distribution of lens
galaxy magnitudes is not the same as the galaxy luminosity function.
Nevertheless, it is straightforward to predict the magnitude
distribution for lens galaxies \citep[e.g.,][]{Fukugita1991}.  For a
singular isothermal sphere lens model, the lensing probability
$F\propto \sigma_{l}^4$, where $\sigma_l$ is the velocity dispersion
of the lensing galaxy \citep{Fukugita1991}.  Assuming a Faber-Jackson
relation $L\propto \sigma^{\gamma_{\rm FJ}}$ \citep{Faber1976}, we
derive a lensing probability $\propto L^{4/\gamma_{\rm FJ}}$.

We begin with a Schechter luminosity function \citep{Schechter1976},
\begin{equation}
\phi_l(L) \, \mathrm{d}L =
\frac{n_*}{L_*}\left(\frac{L}{L_*}\right)^\alpha e^{-L/L_*}\,\mathrm{d}L\, .
\end{equation} 
Factoring in the lensing probability, the
probability for a lens galaxy to have a luminosity $L$ is
\begin{equation}
g_l(L) \, \mathrm{d}L =
\frac{n^\prime_*}{L_*}\left(\frac{L}{L_*}\right)^{\alpha+4/\gamma_{\rm FJ}}e^{-L/L_*}\,\mathrm{d}L\,.
\end{equation} 
We integrate this equation and normalize to a limiting luminosity
$L_\mathrm{lim}$ (thereby accounting for bias due to magnitude-limited
surveys and non-converging functions) to get the cumulative
probability for lens galaxy luminosities greater than $L/L_*$,
\begin{equation}
G_l(L) =
\frac{1-\Gamma(1+\alpha+4/\gamma_{\rm FJ},L/L_*)}
{1-\Gamma(1+\alpha+4/\gamma_{\rm FJ},L_\mathrm{lim}/L_*)}\, ,
\label{eqn.cumfrac}
\end{equation}
where $\Gamma$ is the normalized incomplete gamma function.  This
equation is readily modified to $I$ magnitudes using $L/L_* =
10^{-0.4(I-I_*)}$.  Note that Eq.~\ref{eqn.cumfrac} is fully
degenerate in $\alpha$ and $\gamma_{\rm FJ}$.

We compare two samples to this theoretical prediction.  First, we use
the nine lenses from the present sample that have early-type lens
galaxies (we exclude the two spiral lenses B1600 and PMN2004 and the
two-plane lens B2114).  Second, we note that as this simple
theoretical prediction does not explicitly involve environment, we can
also compare it with the larger sample of early-type lens galaxies
defined by \citet{Rusin2003}, who do not determine the environments of
their lenses.  The Rusin sample spans a larger redshift range than
ours, so we use only the low-redshift systems ($z_{\rm lens} < 0.6$)
with spectroscopic redshifts, a total of 12 lenses.  Five of the nine
lenses in our sample overlap with the Rusin sample. 

We take observed F814W or F791W magnitudes from \citet{Rusin2003} or
from the CASTLES web database,\footnote{In most cases the lens galaxy
magnitude cannot be determined from our data, because the lens galaxy
and quasar images are unresolved.} convert to Cousins $I$, correct for
Galactic extinction, and calculate $I-I_*$.  For $I_*$ we again use
our photometric models of early-type galaxies evaluated at the RS
photometric redshift (if an RS exists) or at the lens spectroscopic
redshift.  For our lens sample, we use the photometric RS redshifts to
insure consistency with our ongoing, larger lens sample, in which many
lens spectroscopic redshifts will not be known.  The use of RS
photometric redshifts instead of spectroscopic redshifts does not
affect any of our conclusions here.

\begin{figure}
\plotone{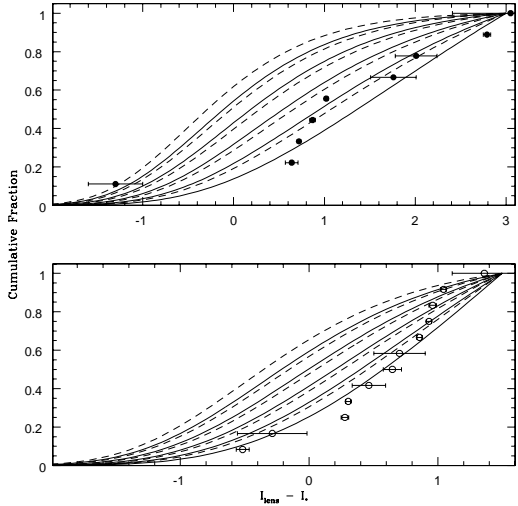}
\caption{Cumulative fraction of lens galaxy magnitudes compared with
  predicted distributions from Eqn.~\ref{eqn.cumfrac}.  Solid curves
  represent $\gamma_{\rm FJ} = 4$; dashed curves are for $\gamma_{\rm
  FJ} = 3.3$, as calculated in \citet{Rusin2003}.  From left to right,
  curves are for faint-end luminosity function slopes of $\alpha =
  -0.6\,{\rm to}\,-1.8$ in intervals of 0.3, with the middle value of
  $-1.2$ shown in bold.  Top panel: Lens galaxies in this paper's
  sample (solid points), with theoretical curves normalized to
  $I-I_*=3$. Bottom panel: $\zlens\leq 0.6$ lens sample of
  \citet{Rusin2003} (open circles).  The lens galaxy magnitude
  distribution in both samples is well-described by the theoretical
  models and reasonable values of $\gamma_{\rm FJ}$ and $\alpha$.
  \label{fig.cumfrac}} 
\end{figure}

The cumulative distributions of observed lens galaxy magnitudes are
shown in Figure \ref{fig.cumfrac}, along with predicted distributions
for different values of $\alpha$ and $\gamma_{\rm FJ}$.  Because our
sample extends fainter than the Rusin sample ($I_{\rm lim} - I_*
\approx 3$ versus 1.5, respectively), the two observed samples are
plotted separately.  There is, however, no significant difference
between our sample and the Rusin low-redshift subsample: the
Kolmogorov-Smirnov (KS) probability is $P=0.42$.  Both samples
indicate that the median low-redshift, early-type lens galaxy has a
magnitude $I - I_* \approx 0.75$, or a luminosity $L \approx 0.5 L_*$.
This median value is directly dependent on the value of $I_*$; as
discussed in Appendix A, our normalization of $I_*$
is brighter by $\sim 1$ magnitude than most modern surveys, so the
median lens luminosity may well be closer to $L_*$. Even allowing for
this, fully half of the elliptical lens galaxies are fainter than
$L_*$, some significantly so (in particular, BRI0952 and MG0751 are 2
to 3 magnitudes fainter than $I_*$, having $L \sim 0.1L_*$).

The figure shows that there are values of $\alpha + 4/\gamma_{\rm FJ}$
leading to theoretical predictions that match the data reasonably well.
High values of $\alpha + 4/\gamma_{\rm FJ}$ are clearly inconsistent with
the data: KS tests rule out $\alpha + 4/\gamma_{\rm FJ} \gtrsim -0.2$ at
the 2$\sigma$ level ($P \le 0.05$).  For reasonable Faber-Jackson slopes
($\gamma_{\rm FJ} \sim 3$ to 4), this would suggest that the faint-end
luminosity function slope is steep, $\alpha \lesssim -1.2$.  We note that
any selection bias against small image separation lenses will tend to
lessen the slope of the observed distribution in Figure 9, as such a bias
will be more pronounced in the less massive (and, by assumption, fainter)
lenses.  This potential bias, the degeneracy between $\alpha$ and
$\gamma_{\rm FJ}$, and the uncertainty in the normalization in $I_*$
prevent us from drawing detailed conclusions here.
 
There are several important results from this exercise.  On the
observational side, the median lens galaxy luminosity is near or below
$L_*$, and the luminosity distribution is very broad.  In other words,
lens galaxies are not all massive, super-$L_*$ ellipticals; rather,
they span a range of luminosities (and, presumably, masses), extending
down at least to $\sim 0.1 L_*$.  On the theoretical side, the
observed luminosity distribution is consistent with simple models of a
Schechter luminosity function weighted by the lensing probability, for
reasonable values of $\alpha$ and $\gamma_{\rm FJ}$.  The key
conceptual point is that, while lensing selects galaxies by mass,
dwarf galaxies are sufficiently common to comprise a moderate fraction
of the lens galaxy sample.  

\subsection{Properties of the Brightest Red Sequence Galaxies\label{sec.brsg}}

In the local universe, X-ray luminous groups of galaxies with
$\sigma\gtrsim 300\,{\rm km\,s^{-1}}$ contain a giant elliptical, the
\emph{Brightest Group Galaxy} or BGG \citep{Mulchaey1998}, which
occupies a unique position in the kinematic and projected spatial
center of the group \citep{Zabludoff1998,van den Bosch2005}.  In this
section we examine the BGGs in our group sample in order to determine
whether any evolution in their properties can be observed.  In two of
three distant, high-$\sigma$ ($\sim$300--500 km s$^{-1}$) groups,
\citet{Momcheva2005} found that the lack of a significant offset of
the BGG from the group centroid is similar to that observed in nearby
groups.  Given the larger sample offered by this photometric study, we
identify the BGG in each candidate group and examine its relation to
the group centroid.  As BGGs at low redshift are typically
ellipticals, we assume here that the brightest galaxy on the selected
RS (BRSG, for brightest red sequence galaxy) is the BGG.

The properties of the BRSG in each group are given in
Table~\ref{tab.brsg}.  Only two lens galaxies, RXJ1131 and MG1654, are
BRSGs.  We also note that two BRSGs, those in B2114 Group 1 and HE2149
Group 4, are known not to be the BGG as their redshifts are
inconsistent with group membership.  The fraction of lens galaxies
that are BRSGs is 0.25 (two lenses out of eight lens groups);
\citet{Momcheva2005} find a BGG fraction of 0.33 (two lenses out of
six lens groups).  These values compare favorably with the suggestion
by \citet{Oguri2005} than $\lesssim 50\%$ of lens galaxies occupy the
dominant halo, the remainder being ``satellite'' galaxies, or
sub-halos within the overall group halo. This result is also not
surprising in light of our conclusion in \S\ref{sec.lensmagdist} that
the lens galaxy is typically not a very luminous galaxy.

\begin{deluxetable*}{lrccccccl}
\tablewidth{0pt}
\tabletypesize{\footnotesize}
\tablecolumns{8}
\tablecaption{Properties of brightest red sequence galaxies\label{tab.brsg}}
\tablehead{\colhead{RS ID} & \colhead{ID$_{\rm BRSG}$} & 
\colhead{$I_{0{\rm ,BRSG}}$} & \multicolumn{2}{c}{Offset from lens} & 
\multicolumn{2}{c}{Offset from centroid} & ID$_{\rm
  BGG}$\tablenotemark{a} & \colhead{Notes} \\   \cline{4-5} \cline{6-7}
 & & & \colhead{(\arcsec)} & \colhead{(kpc)} & \colhead{(\arcsec)} & 
\colhead{(kpc)} &  & }
\startdata
B0712: GROUP 1   &  8471 & 18.387 & \nodata & \nodata & $  28\pm 10$ & $190\pm  70$ & \nodata & \\
MG0751: GROUP 1  &  8626 & 17.824 &   6     &  30     & $  96\pm 18$ & $460\pm  90$ & 8626 & a.k.a. G1 \citep{Tonry1999}\\
MG0751: GROUP 2  &  8387 & 18.171 & \nodata & \nodata & $  34\pm 12$ & $200\pm  70$ & 8476 & Brighter than \citet{Momcheva2005} BGG\\
FBQS0951: GROUP 1 &  7859 & 17.713 & \nodata & \nodata & $ 128\pm 49$ & $350\pm 130$ & \nodata & \\
FBQS0951: GROUP 2 &  9345 & 16.941 & 110     & 450     & $  65\pm 22$ & $270\pm  90$ & \nodata & \\
FBQS0951: GROUP 3 &  8679 & 17.665 & \nodata & \nodata & $  18\pm 20$ & $100\pm 110$ & \nodata & \\
BRI0952: GROUP 1 &  9622 & 19.422 &  71     & 360     & $  84\pm 18$ & $430\pm  90$ &  8406   & Brighter than \citet{Momcheva2005} BGG\\
PG1115: GROUP 3  & 11679 & 18.804 & \nodata & \nodata & $ 106\pm 21$ & $430\pm  90$ & \nodata & \\
PG1115: GROUP 1  & 12600 & 17.824 &  24     & 110     & $  42\pm 19$ & $190\pm  90$ & 12600   & a.k.a. G1 \citep{Impey1998}\\
PG1115: GROUP 2  & 13804 & 17.732 & \nodata & \nodata & $ 101\pm 17$ & $550\pm  90$ & 13764   & Brighter than \citet{Momcheva2005} BGG\\
RXJ1131: GROUP 1 &  7236 & 15.135 & \nodata & \nodata & $  44\pm 22$ & $140\pm  70$ & \nodata & $z=0.104$ from LCRS \\ 
RXJ1131: GROUP 2 &  6323 & 16.6   &   0     &   0     & $  25\pm 22$ & $110\pm 100$ & \nodata & Lens galaxy \\
B1422: GROUP 1   & 13152 & 17.523 &   8     &  40     & $   2\pm 22$ & $ 10\pm 100$ & 13152   & \\
B1600: GROUP 1   &  7950 & 19.714 & \nodata & \nodata & $  88\pm 15$ & $540\pm  90$& \nodata & \\
MG1654: GROUP 1  & 17334 & 17.92  &   0     &   0     & $  78\pm 40$ & $280\pm 140$ & 17334   & Lens galaxy \\
B2114: GROUP 1   &  7943 & 17.841 & 185     & 740     & $ 108\pm 20$ & $430\pm  80$ &  8938   & Not group member \citep{Momcheva2005}\\
HE2149: GROUP 3  &  7568 & 17.663 & \nodata & \nodata & $  80\pm 15$ & $430\pm  80$ & 8364 & Brighter than \citet{Momcheva2005} BGG\\
HE2149: GROUP 4  &  7534 & 19.892 & \nodata & \nodata & $  54\pm 16$ & $360\pm 110$ & 6342 & Not group member \citep{Momcheva2005} \\
\enddata
\tablenotetext{a}{ID of BGG in \citet{Momcheva2005}}
\end{deluxetable*}

\begin{figure}
\plotone{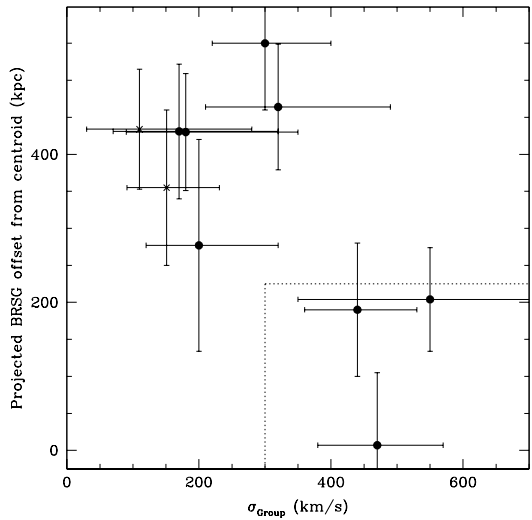}
\caption{Offset of the brightest red-sequence galaxy (BRSG) from the
red sequence centroid as a function of spectroscopic group velocity
dispersion \citep[from][]{Momcheva2005}.  Crosses represent BRSGs
known not to be the brightest group galaxy (see Table \ref{tab.brsg}).
The dotted box indicates the maximum spatial offset of BGGs in groups
with $\sigma_{\rm group}>300\,{\rm km\, s}^{-1}$ in the local Universe
\citep{Zabludoff1998}. As in the local Universe, higher velocity
dispersion groups have a BGG near the group centroid, while lower
velocity dispersion groups groups do not.\label{fig.bgg}}
\end{figure}

Figure \ref{fig.bgg} compares the offset of each BRSG from the raw
projected spatial RS centroid to the measured group velocity
dispersion from \citet{Momcheva2005}.  With one exception (MG0751),
groups with $\sigma\gtrsim 300\,{\rm km \, s}^{-1}$ have a BRSG near
the group centroid, while none of the smaller-$\sigma$ groups have a
central BRSG.  A Spearman rank correlation of $\sigma$ and the
centroid-BRSG offset also suggests a trend (the probability of no
correlation is 0.14).  Our data therefore intimate that there are
dynamically evolved (i.e., virialized) groups with high $\sigma$ and a
central, dominant early-type member by $z\sim 0.5$, groups whose
properties are consistent with evolved (virialized) groups in the
nearby Universe.

\subsection{The Ratio of Quad to Double Lenses}

One unsolved puzzle in lensing studies is why there are so many
quadruple-image lenses compared with double-image lenses
\citep{King1996,Kochanek1996,Keeton1997,Rusin2001,Cohn2004,Keeton2004}.
Highly-flattened lens galaxies are one way to obtain the seemingly
high numbers of quad lenses; another possibility is the presence of a
large tidal shear from mass near the lens galaxy or along the
line-of-sight.  In this latter scenario, quad lenses might be expected
to be located preferentially in complex environments or along
cluttered lines of sight.  Indeed, \citet{Momcheva2005} note that in a
sample of eight lenses, both of the quad lenses and the one quad/ring
lens are in relatively massive groups, while another ring and two of
the doubles are in poorer groups, and the other two doubles have no
groups around the lens galaxies.

Photometrically, we find some weak evidence that quad lenses may have
more complex environments than double lenses. Three of the four quad
lenses and all three ``other'' lenses have associated RSs, while only
two of the five double-image lenses have RSs. However, this difference
is not statistically significant; larger samples are needed to make
any definitive statement.  The lack of strong evidence for a
difference between quad and double lens environments contrasts with
our findings in \citet{Momcheva2005}.  The difference may simply
reflect small number statistics. A more likely explanation is that the
results in M06 were based on detailed calculations of the convergence
and shear due to each lens environment --- calculations we cannot make
in this paper because we do not know the true memberships and velocity
dispersions for all of our groups.  Our complete lens sample will
enable us to reexamine any connections between image number and
environment that may be hidden in the noise of our current sample.

We find no correlation between the number of lensed images and the
presence/absence of line-of-sight structures.  Three of the four
quads, three of the five doubles, and two of the three ``others'' have
at least one line-of-sight structure identified photometrically.
Again we need more data to draw a strong conclusion about any possible
relation.

\section{Conclusions}

We have presented the first results from an on-going wide-field imaging
survey of $\sim 80$ strong gravitational lens systems.  We use a
modified version of the \citet{Gladders2000} group-finding algorithm
to locate potential red sequences, both at the lens redshift and along
the line-of-sight, in the fields of twelve lenses.  This modified
algorithm is capable of detecting $\sigma \gtrsim 200-300$ km s$^{-1}$
poor groups out to $z\approx 0.4$, and higher-$\sigma$ groups out to
higher redshifts with a low rate of false positives and reasonable
completeness.

We draw the following conclusions:

\begin{itemize}

\item Most gravitational lenses lie in complex environments.  For this
  sample, 67\% of lens galaxies (eight of twelve) lie in galaxy
  groups.  As there are no strong biases in this sample, is is
  unlikely this fraction would drop below $\sim 50\%$ once the
  environments of all lenses are surveyed.
\item Most gravitational lenses have interloping structures.  In
  particular, we detect ten groups projected within $\sim 1\farcm5$ of
  seven lenses; i.e., $\approx 60\%$ of the lenses in this sample have
  at least one interloping group that could impact the lens model (see
  also Momcheva et al.~2006).
\item The centroid positions of the lens
  groups are consistent with the directions of the external shears required
  by lens models, for the six systems where we can make the comparison.
  This suggests that lens environments can explain the model shears ---
  which has often been assumed, but not well proven.  In at least one system
  (MG0751), the agreement is seen only when the full virial extent of the
  group is included.
\item The typical gravitational lens is not a super-$L_*$ elliptical.
  Rather, the observed distribution of lens magnitudes is
  well-described by a convolution of lensing probabilities with a
  \citet{Schechter1976} luminosity function, with a median lens galaxy
  luminosity $\sim L_*$ and with known lens galaxies as faint as $\sim
  0.1L_*$.
\item The typical lens galaxy is not the brightest group galaxy.  In
  the current sample, only two of eight lens galaxies (25\%) in a
  group are the brightest group galaxy.
\item As in the local Universe, higher velocity dispersion
  ($\sigma\gtrsim 300 \, {\rm km\, s}^{-1}$), intermediate-redshift
  galaxy groups have a brightest group galaxy near the group centroid,
  whereas the brightest group galaxy typically lies outside the group
  center in lower velocity dispersion groups.  This result suggests
  that the higher $\sigma$ groups are more dynamically evolved 
 (i.e., virialized) than the
  lower-$\sigma$ groups and that evolved (virialized) groups exist by $z\sim
  0.5$.
\item In total, we report the photometric discovery or recovery of 18
  candidate poor groups of galaxies in the redshift range $z\sim 0.2$
  to 0.6.  The average red sequence slope is $0.076\pm 0.025$, in
  agreement with that of nearby groups.  The measured scatter about
  the red sequence is comparable to that expected from random photometric
  errors. When analysis of the full lens sample is complete, we expect
  to find $\sim$100 groups at these intermediate redshifts, permitting
  detailed studies of the evolution of group structure and members.
\end{itemize}

\begin{figure*}
\plottwo{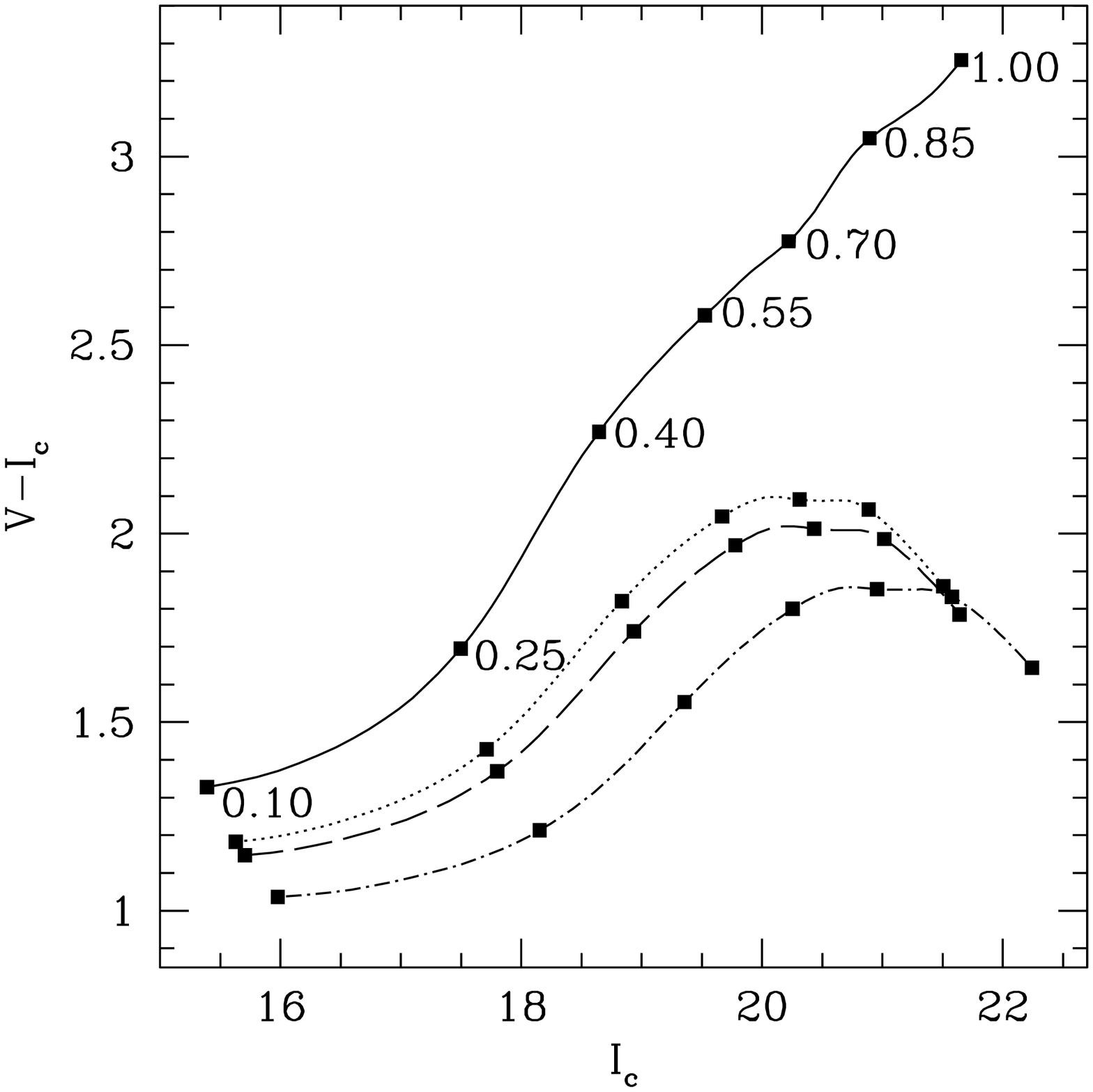}{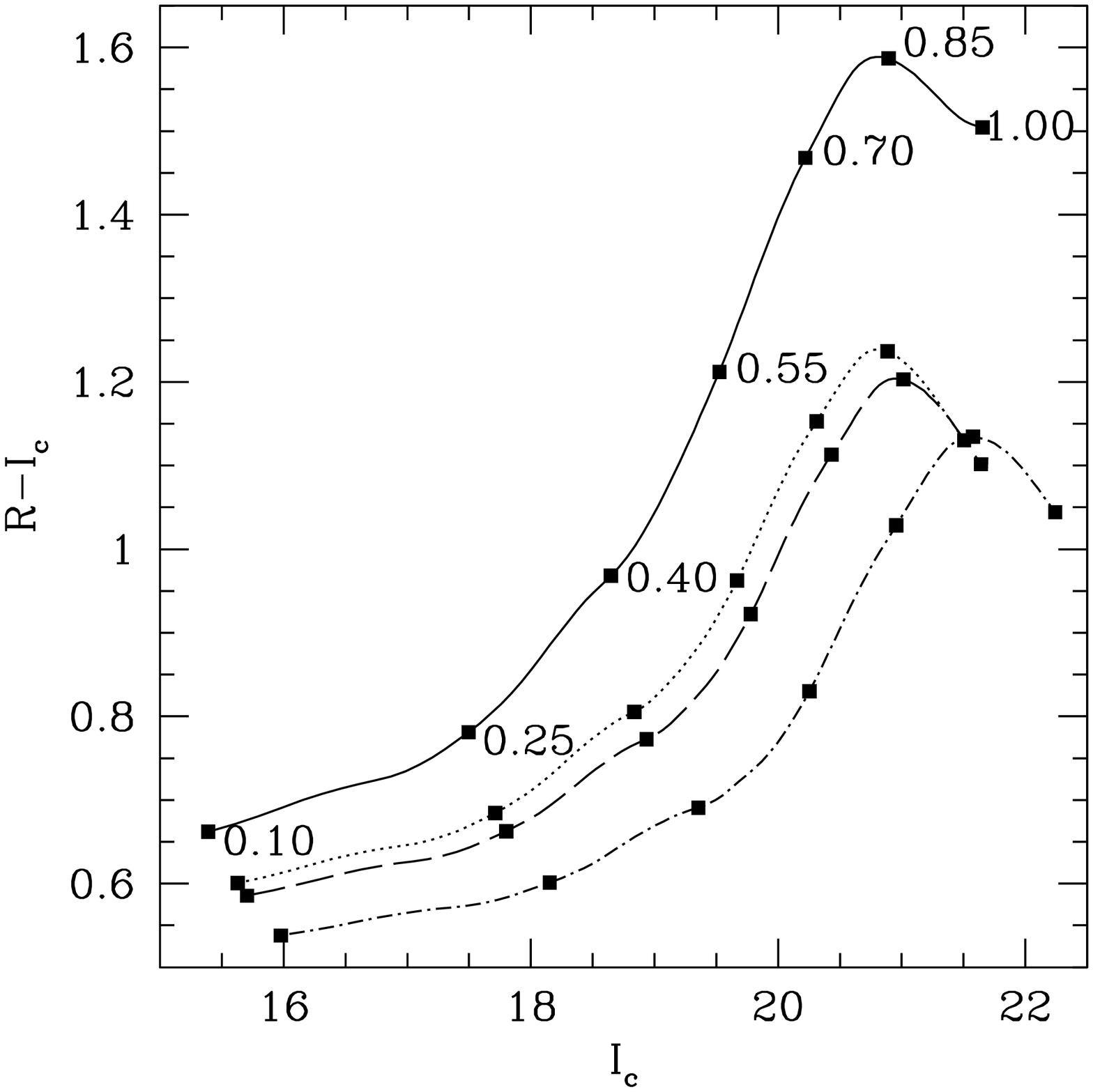}
\caption{Model galaxy $V-I$ (left) and $R-I$ (right) colors versus $I$
magnitude for an $L_*$ galaxy.  Curves indicate E/S0 (solid), Sa
(dotted), Sb (dashed), and Sc (dot-dashed) galaxies and range from
$z=0.1\, -\, 1.0$.  Boxes mark redshift intervals of $\Delta z=0.15$,
with the corresponding numerical redshift indicated along the E/S0
track.\label{fig.model_vi} }
\end{figure*}

\begin{acknowledgements}

The authors are grateful for support for this project in the form of
National Science Foundation grant AST-0206084 and NASA LTSA award
NAGS-11108.  CRK was supported in part by NASA through Hubble
Fellowship grant HST-HF-01141.01-A from the Space Telescope Science
Institute, which is operated by the Association of Universities for
Research in Astronomy, Inc., under NASA contract NAS5-26555. IM
acknowledges the support of the Martin F. McCarthy Scholarship in
Astrophysics awarded by the Vatican Observatory.

We thank Dennis Zaritsky, Greg Rudnick, Luc Simard, 
and Chien Peng for helpful discussions along the way.  We thank
the referee, Chris Kochanek, for comments that improved
this paper.  We also thank
Mike Brown (the Aussie) and Doug Clowe for snippets of software
used in the analysis.  We especially thank Emilio Falco for his vital
assistance in the formulating the initial target lists.

Observations were made at Kitt Peak National Observatory and the
Cerro-Tololo Inter-American Observatory, as part of the National
Optical Astronomy Observatory, which is operated by the Association of
Universities for Research in Astronomy, Inc. (AURA) under cooperative
agreement with the National Science Foundation.  The authors also
would like to express their appreciation to the excellent staff at
both of these facilities, especially to the excellent telescope
operators and the computer support staff who saved many gigabytes of
data on more than one occasion.

The authors wish to recognize and acknowledge the significant
cultural role and reverence that Kitt Peak holds within the Native
American community, in particular the Tohono O'odham Nation.  We are
most fortunate to have the opportunity to conduct observations from
this mountain.

\end{acknowledgements}


\appendix
\section{Predicted Galaxy Photometric Properties
\label{sec.pred_phot}} 

For this paper, we determine the photometric properties of galaxies as
a function of redshift from simple evolutionary models already used
for analysis of lens galaxies \citep{Keeton1998}.  These models begin
with spectrophotometric models by \citet{Bruzual1993}, assuming solar
metallicity and a star formation epoch starting at $z=5$.  We assume
that early-type galaxies are formed by a 1-Gyr burst of star formation
with a \citet{Salpeter1955} initial mass function (IMF) followed by
passive evolution to the present day.  We model late-type galaxies
with a \citet{Scalo1986} IMF and with the star formation rates from
\citet{Guiderdoni1988} for Sa, Sb, and Sc galaxies.  We use these
models to compute the color-, $K$-, and evolutionary corrections
needed to predict the apparent magnitude and colors of an $L_*$ galaxy
as a function of redshift.  For this work, we define $L_*$ to be the
luminosity of a $z=0$ galaxy with an absolute $B$-band magnitude of
$M_B(L_*) = -19.9+5\log h$. The model color-magnitude curves are shown
in Figure~\ref{fig.model_vi}.

We also consider the impact of different normalizations of $M_B(L_*)$
on our results. We have converted each of the published normalizations
to a consistent cosmology and to $M_B$, the latter by using the $z=0$
colors for elliptical galaxies calculated by \citet{Fukugita1995}.
Recent analysis of the density-dependence of the galaxy luminosity
function from the 2dF Galaxy Redshift Survey \citep{Colless2001} 
is presented in \citet{Croton2005}, who find
$M_B(L_*)$ ranging from $-18.3 + 5\log h$ for early-type galaxies in
voids to $-19.8 + 5\log h$ for early-type galaxies in clusters, with a
value $\approx -19.4 + 5\log h$ for the entire volume.
\citet{Bernardi2005} find a value of $M_B(L_*)=-18.9 + 5\log h$ for
early-type galaxies from the Sloan Digital Sky Survey.  Other galaxy
surveys have values of $M_B(L_*)$ ranging from $\approx -18.7+ 5\log h$ to
$\approx -19.4+ 5\log h$ \citep[e.g., see Table 1 in][]{Gonzalez2000}. 

Analytically, the change in the color $V-I$ of a galaxy of magnitude $I$ 
on the red sequence from a change in
normalization of $I_*$ of $\Delta I_*$ is:
\begin{equation}
\Delta (V-I) = \beta_V \Delta I_*\, ,
\end{equation}
where $\beta_V$ is the slope of the red sequence in $V-I$.  Taking the
faintest value of $M_B(L_*)= -18.3 + 5\log h$ from the surveys above
and $\beta_V=0.08$ (\S\ref{sec.algorithm}),
the maximum change in the color of a red sequence galaxy is $\Delta
(V-I)= -0.128$.  Over the redshift range of $0.2 < z < 1.0$, the
average change in photometric redshift as a function of change in
color is $\Delta z / \Delta (V-I) = 0.133$, so the maximum change in
redshift from a change in normalization would be $\Delta z \approx
-0.02$. We test this calculation by running the group-finding
algorithm on several groups with $M_B(L_*)=-18.3 + 5\log h$, and
indeed the detected group redshifts are lower by $\approx
0.02$.  We therefore note that our photometric group redshifts may be
systematically biased toward higher redshifts by as much as $\Delta
z=0.02$, smaller than the measured scatter between groups photometric
and spectroscopic redshifts given in \S\ref{spec.comp}.

\end{document}